\documentclass[journal=jacsat,manuscript=article,layout=twocolumn]{achemso}

\usepackage{amsthm}
\usepackage{amsmath}
\usepackage{bm}
\usepackage{color}
\usepackage{gensymb}

\usepackage{accents}
\newcommand{\dbtilde}[1]{\accentset{\approx}{#1}}

\graphicspath{{./fig/}}

\newcommand{\EQ}{\begin{equation}}
\newcommand{\EN}{\end{equation}}
\newcommand{\EQA}{\begin{eqnarray}}
\newcommand{\ENA}{\end{eqnarray}}

\newcommand{\Eq}[1]{Eq.~(\ref{#1})}

\newcommand{\App}[1]{Appendix~\ref{#1}}
\newcommand{\Eqs}[2]{Eqs.~(\ref{#1}) and~(\ref{#2})}

\newcommand{\Fig}[1]{Fig.~\ref{#1}}
\newcommand{\Tab}[1]{Table~\ref{#1}}
\newcommand{\Figs}[2]{Figs.~\ref{#1} and \ref{#2}}

\newcommand{\Nu}{\mbox{\rm Nu}}
\newcommand{\Kn}{\mbox{\rm Kn}}

\newcommand{\SSSS}{\mbox{\boldmath ${\sf S}$} {}}
\newcommand{\ttau}{\mbox{\boldmath ${\tau}$} {}}

\newcommand{\nsw}{n^{\rm sw}}
\newcommand{\dotrho}{\dot{\rho}}
\newcommand{\tildensw}{\tilde{n}^{\rm sw}}

\newcommand{\uu}{\bm{u}}
\newcommand{\vvv}{\bm{v}}
\newcommand{\aaa}{\bm{a}}

\newcommand{\qq}{\bm{q}}
\newcommand{\xx}{\bm{x}}
\newcommand{\JJ}{\bm{J}}
\newcommand{\UU}{\bm{U}}
\newcommand{\VV}{\bm{V}}
\newcommand{\nhatp}{\hat{\bm{n}}_p}
\newcommand{\nm}{\,{\rm nm}}
\newcommand{\K}{\,{\rm K}}

\newcommand{\Ox}{{\rm O}}
\newcommand{\NOx}{{\rm NO_x}}
\newcommand{\CO}{{\rm CO}}
\newcommand{\COO}{{\rm CO_2}}
\newcommand{\OO}{{\rm O_2}}
\newcommand{\NN}{{\rm N_2}}
\newcommand{\SiO}{{\rm SiO}}
\newcommand{\SiOO}{{\rm SiO_2}}
\newcommand{\HHO}{{\rm H_2O}}
\newcommand{\OH}{{\rm OH}}
\newcommand{\Hy}{{\rm H}}
\newcommand{\M}{{\rm M}}
\newcommand{\ms}{{\rm ms}}
\newcommand{\nab}{\bm{\nabla}}
\newcommand{\cs}{c_{\rm s}}

\SectionNumbersOn

\title{Silicon-monoxide flames: the nucleation and condensation of silica fume}

\author{Nils Erland L. Haugen}
\affiliation{SINTEF Energy Research, Trondheim, Norway}
\affiliation{Division of Energy Science, Lule{\aa} University of Technology, Lule{\aa}, Sweden}
\email{Nils.E.Haugen@sintef.no}

\author{Axel Brandenburg}
\affiliation{Nordita, KTH Royal Institute of Technology and Stockholm University, Hannes Alfv\'ens v\"ag 12, SE-10691 Stockholm, Sweden}
\alsoaffiliation{The Oskar Klein Centre, Department of Astronomy, Stockholm University, AlbaNova, SE-10691 Stockholm, Sweden}
\alsoaffiliation{McWilliams Center for Cosmology \& Department of Physics, Carnegie Mellon University, Pittsburgh, PA 15213, USA}
\alsoaffiliation{School of Natural Sciences and Medicine, Ilia State University, 3-5 Cholokashvili Avenue, 0194 Tbilisi, Georgia}
\author{Bernd Friede}
\affiliation{Elkem ASA, Kristiansand, Norway}
\author{Rolf G. Birkeland}
\affiliation{Elkem ASA, Kristiansand, Norway}

\begin{document}

\maketitle

\begin{abstract}
Silica fume is a valuable by-product from the silicon and ferrosilicon production. It is therefore important to understand the impact on the silica fume quality when converting the furnace feed from fossil-based to renewable reduction materials.
Using self-consistent numerical simulations of the nucleation and condensation process, we present a detailed study of the silica fume formation process. 

It is found that the most critical physical effect that determines the final particle size distribution is coalescence due to Brownian motion.
Furthermore, it is crucial to use appropriate thermophysical parameters in order to reproduce reliable particle size distributions.
Contrary to what has been done in previous studies on the same topic,
this is now done by using reasonable expressions for surface energy, saturation pressure and the nucleation pre-exponential factor. 

It is also found that under conditions relevant to furnaces, the liberation of latent heat leads to
an explosive chain reaction of particle nucleation and condensation when the first particles nucleate and start growing due to condensation.
This process continues until the relative saturation pressure of silicon dioxide is reduced to unity.

Finally, it is found that the Lagrangian approach for particle tracking is more flexible and accurate, and also more CPU efficient, than the Eulerian approach.
\end{abstract}

\section{Introduction}

Refining natural metal oxide ores into metals requires a reducing agent to remove oxygen from the ore.
Some of the most common reducing agents used by the metallurgic industry today are fossil carbon-based materials such as coal or coke.
Utilization of fossil materials in reduction processes results in large global CO$_2$ emissions.
There is therefore a drive to substitute fossil-based reducing agents with renewable materials, such as, e.g., bio-carbon.

When producing silicon or ferrosilicon, silica fume particles are a major by-product.
Up until the 1980s, the smoke consisting of silica fume particles from this production was an environmentally challenging side stream,
but since then silica fume has been turned into a valuable commodity due to its use in various advanced materials.
It is particularly valuable because of its use as pozzolan in concrete, and its ability to significantly increase particle packing in refractory materials \cite{Friede2011}.
Silica fume coming from silicon and ferrosilicon furnaces consists of non-porous, spherical nanoparticles of almost pure amorphous silicon dioxide \cite{Friede2018}.
When converting these furnaces process from fossil-based to bio-based reducing agents, it is crucial that the quality of the silica fume is not degraded.
It is therefore important to fully understand the mechanisms by which the silica fume particles are formed.
Such knowledge can then be used to optimize the production process in order to maintain a stream of high quality silica fume also for bio-based reducing agents.
In particular, it is important to (i) maintain a high purity material with negligible amounts of contamination from the bio-based reducing agent, and (ii) have approximately the same physico-chemical properties, such as pozzolanic activity and particle-size distribution of silica fume as with fossil based reducing agents.

The gaseous silicon-dioxide that will eventually form the 
amorphous silica fume particles is produced by 
combustion of silicon-monoxide ($\SiO$) that emerges from the furnace charge together with carbon-monoxide ($\CO$) \cite{Schei1998}.
When the mixture of $\SiO$ and $\CO$ meet the air that flows into the furnace above the charge, it burns in an intense flame right above the charge due to the high oxygen affinity of silicon. 

The silicon atom in molecular SiO2 has a coordination number of 2, and is therefore in a high energy, unstable state since its preferred coordination number is 4 (SiO$_4^{4-}$), as it can be seen in most crystalline and amorphous silica polymorphs and silicates. 

The combustion products are carbon-dioxide ($\COO$) and silicon-dioxide ($\SiOO$): 
\EQ
\SiO+x\CO+\frac{1}{2}(1+x)\Ox_2\to\SiOO+x\COO,
\EN
where $x$ denotes the fraction of $\CO$ over $\SiO$.

$\SiOO$ molecules will start to nucleate into $\SiOO$ particles in the post-flame region.
Molecular $\SiOO$ that is still in the gas phase will then condensate on these nuclei, or 
generate new nuclei.
Already in 1971, Ulrich studied the formation of pyrogenic silica, also know as fumed silica \cite{Ulrich1971}.
He found that the minimum nucleation radius is smaller than the size of a single molecule, which contradicts the findings of later authors.
The results of Ulrich  \cite{Ulrich1971} are therefore not directly useful for this study.
More recently, in 2020, Gonzalez-Farina made a numerical study of the formation of silica fume
in her PhD thesis \cite{Gonzalez-Farina2020_PhD} and also in a separate journal paper \cite{gonzalez2020}.
Finally, in 2022, Vachaparambil et al. \cite{Vachaparambil2022} used OpenFoam
to investigate the formation of silica fume in a simplified two-dimensional geometry of the furnace hood.
To the best of our knowledge the above-mentioned works are the only publications available in the open literature that use numerical simulations to study the evolution of silica fume from the silicon and ferro-silicon industries.
Although some work has been done related to the formation of silica fume, there is limited literature available that describe in detail how silica fume is formed \cite{Andersen2023}.

Due to more stringent restrictions on emissions of nitrogen-oxides (NO$_x$) \cite{Cusano},
since the early 2010s several papers and PhD theses have used numerical simulations to study how the combustion of $\SiO$ influences $\NOx$ emissions
from silicon producing furnaces 
\cite{Panjwani2011,Kamfjord2012,Olsen2012,Myrhaug2012,Panjwani2013,Brede2013,Ness2013}.
These works did not, however, consider silica fume. Furthermore, in the work of Sloman et al. \cite{Sloman2017},
the authors studied the heat and mass transfer in a silicon pilot furnace with a focus on the charge zone.

In the current work, the aim is to perform detailed simulations
of $\SiO$/$\CO$ oxidation and the resulting nucleation and condensation of $\SiOO$.
This will be done by incorporating all relevant physical and chemical effects and
by
using the best available values for the relevant parameters. In order to start building a fundamental understanding of the evolution of silica fume particles, the focus here is on homogeneous conditions. This means that the gaseous reactants are considered premixed and that there are no spatial gradients in the domain.
The extension to non-premixed and non-homogeneous conditions will be studied in later works.

\section{Equations}

\subsection{Particle equations}

We begin by comparing two numerical approaches for the particles: the Lagrangian and Eulerian approaches.
The focus is nevertheless on the Lagrangian approach, which will be
presented next, while the Eulerian particle approach is presented in
\App{eulerian_part}.

Due to the enormous number of particles involved in the applications we are interested in, it is not feasible to track every single particle.
Instead we lump similar particles together in what we call swarms (or super-particles).
All the individual particles in a swarm are identical in size,
velocity and temperature, and they are assumed to be homogeneously distributed within the nearest grid
cell with a particle number density of $\nsw$.
Further details regarding our implementation of the swarm or
super-particle approach are given elsewhere \cite{Li2022}.

The evolution equation for each particle position is given by
\EQ
\frac{d\xx_p}{dt}=\vvv_p,
\EN
where $\vvv_p$ is the particle velocity. 
The particle velocity equation reads
\EQ
\frac{d\vvv_p}{dt}=-\aaa_p^{\rm drag}+\aaa_p^{\rm Brown},
\EN
where
\EQ
\aaa_p^{\rm drag}=\frac{\vvv_p-\uu}{\tau_p}
\EN
is the particle acceleration due to drag from the gas phase, $\uu$ is the velocity of the carrier fluid, and
\EQ
\label{eq:brown}
\aaa_p^{\rm Brown}=\nhatp \sqrt{\frac{2}{\tau_p \Delta t} \frac{k_B T}{m_p}}
\EN
is the particle acceleration due to Brownian forces.
A validation of this expression for Brownian motion of particles is found in \App{app:brown}.
For the particles of interest, the particle radius $r_p$ is smaller than the mean free path of the carrier fluid.
The particle response time is therefore given by the Epstein expression:
\EQ
\label{taup}
\tau_p=\frac{r_p \rho_{\ms}}{\cs \rho},
\EN
where $\cs$ is the speed of sound, and $\rho$ and $\rho_{\ms}$ are the material densities of the fluid and the particles, respectively. 
In the expression for the Brownian acceleration, $\Delta t$ is the time-step of the simulation,
$m_p$ is the particle mass and $\nhatp$ is a vector where each element has a zero-mean, unit variance, independent Gaussian random number. 
For numerical stability, the time step must be smaller than the particle response time.
Due to the small radius of the particles of interest here, the response time given by \Eq{taup} becomes very short -- resulting in correspondingly short time steps.
To avoid the problem of too small time steps of the simulation, we assume that the particles have the same velocity as the embedding fluid, except for the Brownian velocity.
The relative velocity difference between a particle and the fluid, $\vvv_{\rm rel}=\vvv_p-\uu$, is then given as 
\EQ
\vvv_{\rm rel}=\aaa^{\rm Brown}\tau_p.
\EN
Hence, the particle velocity becomes
\EQ
\label{eq:vrel}
\vvv_{p}=\uu+\aaa_p^{\rm Brown}\tau_p.
\EN

The radius of the particles can increase through condensation or coagulation, and it can decrease due
to evaporation. Since evaporation is essentially the inverse of condensation, the evolution equation
for the particle radius due to condensation/evaporation is given by 
\EQ
\label{drpdt}
\frac{dr_p}{dt}=A(C_{\SiOO}-C_{\rm sat}) \sqrt{T},
\EN
where $T$ is the temperature of the carrier fluid, $C_{\SiOO}$ is the concentration of gaseous silicon dioxide, $C_{\rm sat}$ is the silicon dioxide saturation concentration, and
\EQ
A=\sqrt{\frac{8k_B}{\pi m_{\SiOO}}}\frac{M_{\SiOO}}{4\rho_{\ms}}
\EN
(see \App{app:condensation} for more detail).

Coagulation of two small particles into one large particle 
is handled in the same way as done previously by Li et al.\ \cite{Li2017,Li2022}, where the collection scheme is that of Shima \cite{Shima2009}.
This approach will now briefly be presented in the following.
Let us consider two swarms (super-particles) $i$ and $j$ within the same grid cell and with particle number densities $\nsw_i$ and $\nsw_j$, respectively.
Furthermore, let us define $i$ and $j$ such that $\nsw_j > \nsw_i$. The collection probability of a particle in swarm $i$ with swarm $j$ within the current time step is then given by
\EQ
p_{ij}=\sigma_{ij}\nsw_j|\vvv_i-\vvv_j|E_{ij} \Delta t,
\EN
when $E_{ij}$ is the collection efficiency, which is here set to unity,
$\sigma_{ij}=\pi (r_i+r_j)^2$ is the collectional cross section, and $r_i$ is the radius of particle $i$.
It is clear that the time step must be so small that the collection probability is always well below unity.
For each swarm pair in a given grid cell, one now draws a random number between zero and unity and compares it with $p_{ij}$.
If the random number is smaller than $p_{ij}$, a collection event is considered to have occurred, otherwise nothing happens.
In the case of a collection event, the mass of particle $j$ is added to the mass of particle $i$, while the mass of particle $j$ is unchanged,
i.e., $\tilde{m}_i=m_i+m_j$ and $\tilde{m}_j=m_j$, where the tilde represents the new value.
Furthermore, the number density of swarm $i$ is subtracted from the number density of swarm $j$, while the number density of swarm $i$ is unchanged, or, in mathematical terms: $\tildensw_i=\nsw_i$ and $\tildensw_j=\nsw_j-\nsw_i$. This framework ensures mass conservation while providing the correct coagulation rate.
Since all particles within the same swarm have the same velocity, particles within a swarm cannot collide with each other.

The equation for the particle temperature $T_p$ is given by
\EQ
\label{temp_part}
\frac{dT_p}{dt}=\frac{q_{\rm rad}+q_{\rm latent}-q_{\rm conv}}{m_p c_\mathrm{p}},
\EN
where $c_\mathrm{p}$ is the heat capacity of the particle,
\EQ
q_{\rm latent}=\frac{\dot{m}_{\rm cond}\Delta H}{M_{\SiOO}}
\EN
represents the release of latent heat due to the condensational phase change from gas to liquid, 
$\Delta H$ is the latent heat of condensation, $M_{\SiOO}$ is the molar mass of silicon dioxide,
\EQ
\label{mdot}
\dot{m}_{{\rm cond}}=4\pi r_p^2 \frac{dr_{p}}{dt} \rho_{\ms}
\EN
is the rate of change of particle mass due to condensation/evaporation,
\EQ
q_{\rm conv}=H_{\rm trans}A_p(T_p-T)
\EN
is the convective heat lost to the gas phase,
$H_{\rm trans}=\Nu \, k_\mathrm{g}/2r_p$ is the heat transfer coefficient, $\Nu$ is the Nusselt number, 
which, in the continuum regime,
can be set to two when the relative velocity between particle and fluid is close to zero \cite{Bird2007}, 
$A_p=4\pi r_p^2$ is the particle surface area, $k_\mathrm{g}$ is the conductivity of the fluid, and 
\EQ
q_{\rm rad}=4\pi r_p^2(T_{\rm wall}^4-T_p^4)\,\sigma_{\rm SB}q_{\rm abs}
\EN
is the radiative heat loss to the surroundings.
Here, $T_{\rm wall}$ represents the temperature of the walls as seen by the particle,
$\sigma_{\rm SB}$ is the Stefan--Boltzmann constant, and $q_{\rm abs}$ is the absorption coefficient.
For a black body emitter with $r_p \gg \lambda$, where $\lambda$ is the wavelength of light, $q_{\rm abs}=1$.
For smaller particles, where $r_p \ll \lambda$, the absorption coefficient is given by \cite{Modest2013}
\EQ
q_{\rm abs}=-\frac{8\pi r_p}{\lambda} \, \mathrm{Im}\,\left( \frac{m^2-1}{m^2+2}  \right),
\EN
where $m=n-\mathrm{i}k$ is the complex index of refraction with $n$ being the refractive index and $k$ the absorptive index \cite{Modest2013}.
It is not entirely clear from the literature what the complex index of refraction for $\SiOO$ particles should be,
but the authors of a recent paper \cite{filmetrics} suggest $n\sim 1.44$ and $k\sim 0$ for high temperatures.
The imaginary part is expected to be small but not zero, and we therefore assume a value of $k = 0.005$.
In the following, we will use $n = 1.44$ and $k = 0.005$.
This means that $\mathrm{Im}\left[(m^2-1)/(m^2+2)\right]=-0.0026$.

\subsubsection{Particle nucleation}
\label{nucleation}

In this work, we do not consider ``external'' nucleation sources, such as, e.g., ash particles.
Instead, nucleation occurs at a rate by which a nucleus with a radius equal to the critical radius is formed.
Here, the critical
radius is given by
\EQ
\label{rmin}
r_{\rm  min}=\frac{2\gamma v_c}{k_BT \ln S_e},
\EN
and the corresponding rate by which this nucleation occurs is given by
\EQ
\label{nucleation_rate}
J=J_0\exp\left(-\frac{16\pi\gamma^3 v_c^2}{3(k_B T)^3(\ln S_e)^2}\right),
\EN
where $J_0$ is the nucleation rate coefficient, $S_e=C_{\SiOO}/C_{\rm sat}$ is the ratio of the
SiO$_2$ gas-phase concentration over its saturation concentration,
$v_c=4.5\times 10^{-29}$~m$^3$ is the volume of a SiO$_2$ molecule,
$k_{\rm B}$ is the Boltzmann constant and
$\gamma$ is the surface energy of a $\SiOO$ particle. 

Instead of adding a new Lagrangian swarm particle at every grid cell for
every time step, we define a fluid variable $y_{\rm nucl}$ that carries the mass of
those nuclei that, within a given volume of fluid, have been generated,
but not yet turned into particles.  When this density exceeds a given
threshold, $\rho y_{\rm nucl}^{\rm thresh}$, we generate a new Lagrangian particle
that contains all these nuclei, while setting the value of $y_{\rm nucl}$ to zero.
The evolution equation for $y_{\rm nucl}$ is
\EQ
\frac{\partial \rho y_{\rm nucl}}{\partial t}+\nabla\cdot (\rho \uu y_{\rm nucl})=\dotrho_{{\rm nucl}},
\EN
where the source term
\EQ
\label{F_nucl}
\dotrho_{\rm nucl}=\frac{4}{3}\pi r_{\rm min}^3 \rho_{\ms}J
\EN
corresponds to a sink in the fluid equations for density and species
mass fractions; see Eqs.~(\ref{species_cont}) and (\ref{species_cond}) due to nucleation.
In non-conservative form, which is the form used in the code, the equation becomes
\EQ
\label{dcdt}
\frac{\partial y_{\rm nucl}}{\partial t}+\uu\cdot\nabla y_{\rm nucl}=
\dot{y}_{\rm diff}
+\frac{\dotrho_{{\rm nucl}}(1+y_{\rm nucl})}{\rho},
\EN
where $\dot{y}_{\rm diff}=D_{\rm s} (\nabla^2 y_{\rm nucl}+ \nabla \ln \rho \cdot \nabla y_{\rm nucl})$ is a diffusion term with
$D_{\rm s}$ being a small diffusion coefficient that is only needed for numerical stability. The $y_{\rm nucl}\dotrho_{{\rm nucl}}/\rho$ term in \Eq{dcdt} comes from the fact that the generation of nuclei, given by $\dotrho_{{\rm nucl}}$, acts as a sink in the continuity equation. 

\subsection{Silica fume particle interactions}

\begin{figure*}[t!]\begin{center}
\includegraphics[width=\textwidth,trim={0cm 4cm 0cm 4cm},clip]{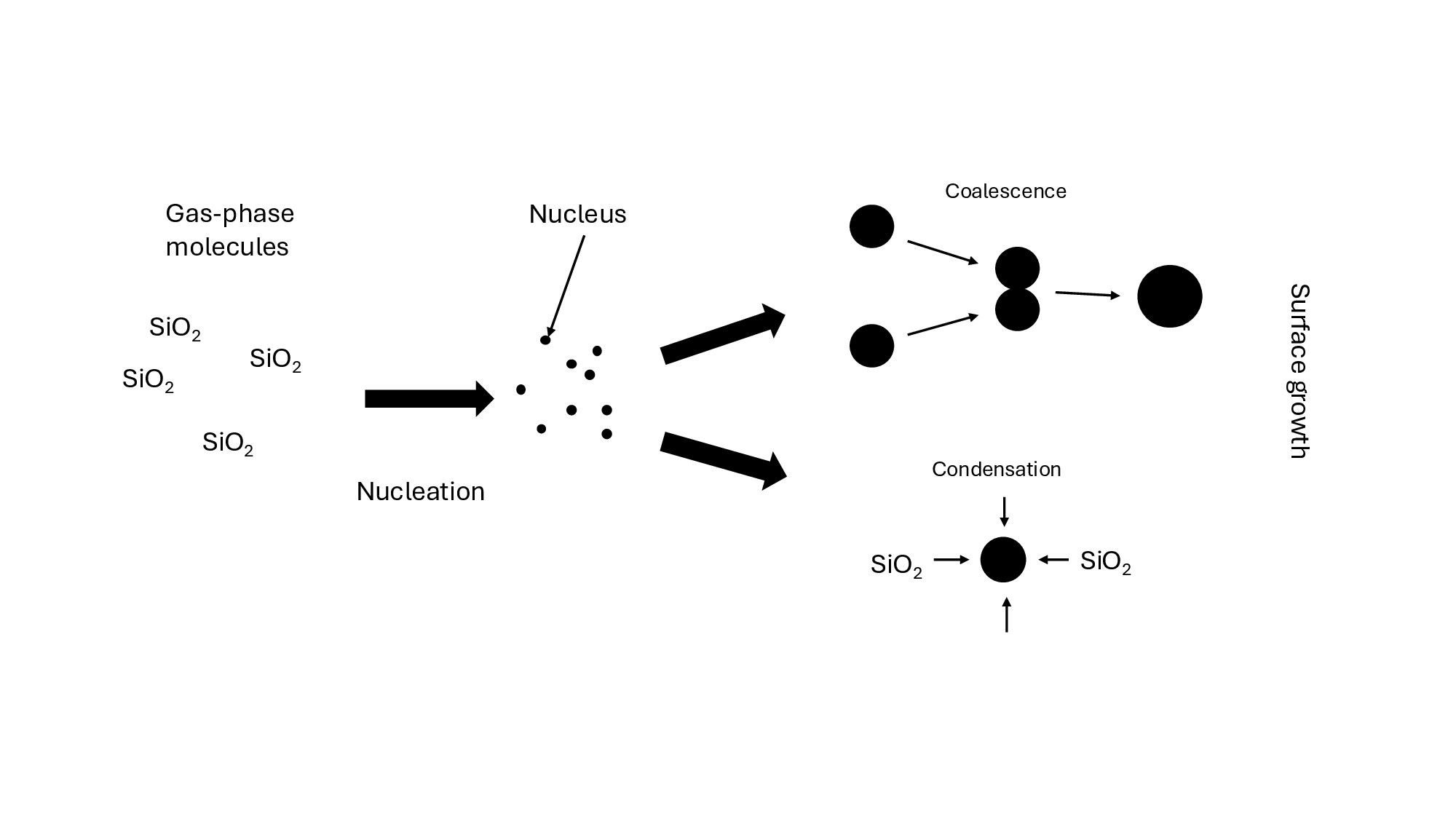}
\end{center}\caption[]{
Schematic view of particle growth from gas phase monomers to silica fume particles through nucleation, condensation and coalescence. (Modified from \citet{Gonzalez-Farina2020_PhD}.)
}\label{ms_interaction_regimes}\end{figure*}

In this section we will postulate a simplified mechanism for how $\SiOO$ molecules grow to the final silica fume particles based on known mechanisms for step polymerization by nucleophilic addition\cite{Billmeyer1984}. 
A schematic view of this is presented in \Fig{ms_interaction_regimes}.
When $\SiOO$ molecules interact, they form a $\SiOO$ polymer if they are in the right temperature range (referred to as nucleation in \Fig{ms_interaction_regimes}).
The polymers grow due to diffusion of $\SiOO$ monomers to their surface (condensation), but also due to collision with other polymers (coalescence).
When two polymers collide, due to energetic reasons they will form one larger spherical polymer if the temperature is above a certain level $T_1$.
If the temperature exceeds yet another higher critical level $T_2$, the chemical bonds between the monomers will break -- and polymers can no longer exist and the monomers transcend into the gas phase.
This process can be considered a sublimation.
The detailed physical chemistry of this process is not well know.
In the following we will therefore apply a droplet approximation for the silica fume.
In this approximation, $T_1$ becomes the melting temperature of silica fume, while $T_2$ is its boiling temperature.
As silica fume does not have a clearly defined melting point, we use the melting point of $\SiOO$ as the lower temperature limit for coalescence.

For temperatures above the boiling point of quartz,
no silica fume particles can exist. Between the melting temperature of 1823 K and the boiling temperature, silica fume particles that collide  will coalesce into one larger particle. Moving to lower temperature, particles with temperatures somewhat below the melting temperature may still stick together and form a neck-shaped connection. If formed, this connection is relatively strong. But, depending on exact temperature and properties of the impact, the two particles may also separate again after the collision.
Finally, at even lower temperatures, particles may form agglomerates that are connected by Van der Waals forces if the collision energy is low enough.
Such agglomerates are, however, typically formed only in piles of silica fume particles behind the filter.

\subsection{Fluid equations}
When $\rho$ is the fluid density, the continuity equation is given by
\EQ
\label{species_cont}
\frac{D \rho}{D t}=-\nabla\cdot\uu -\dotrho_{{\rm cond}}-\dotrho_{{\rm nucl}},
\EN
where $D/Dt=\partial/\partial t+\uu\cdot\nab$ is the advective derivative, and $\dotrho_{{\rm cond}}$ and $\dotrho_{{\rm nucl}}$
are the sinks due to condensation and nucleation, respectively.
Here, $\dotrho_{{\rm nucl}}$ has already been defined in \Eq{F_nucl} while
the sink due to condensation is given by
\EQ
\dotrho_{\rm cond}=\sum_{j=1}^{N_{\rm part,cell}} \dot{m}_{{\rm cond},j} \nsw_j,
\EN
where $\dot{m}_{{\rm cond},j}$ is calculated based on \Eq{mdot}.
Furthermore, the momentum equation for the carrier fluid is given by
\EQ
\frac{D \uu}{D t}=-\frac{1}{\rho}
\left(
\nab P-\nabla\cdot\ttau 
\right).
\EN
In the above equation, $P$ is fluid pressure, $\tau=2\rho\nu\SSSS$ is the stress tensor,
$\nu$ is kinematic viscosity, and $S_{ij}=\frac{1}{2}(\partial u_i/\partial x_j+\partial u_j/\partial x_i) -\frac{1}{3}\delta_{ij}\nabla\cdot\uu$ 
is the traceless rate of strain tensor.

In conservative form, the evolution equation for the mass fraction of species $k$ is given by
\EQ
\frac{\partial \rho Y_k}{\partial t}+\nabla\cdot (\rho \uu Y_k)=-\nabla\cdot \JJ_k+\dot{\omega}_k
-\dotrho_{{\rm cond},j}-\dotrho_{{\rm nucl},j},
\EN
where $\JJ_k=\rho Y_k \VV_k$ is the diffusive flux of species $k$,
$\VV_k$ is the diffusive velocity obtained from the mixture average approximation \cite{Babkovskaia2011, Hirschfelder1969},
and $\dot{\omega}_k$ is the source due to chemical reactions.
Based on this, we obtain the following (non-conservative) equation for the mass fraction of species $k$,
\EQ
\label{species_cond}
\frac{D Y_k}{Dt}=-\nabla\cdot \JJ_k+\dot{\omega}_k+\frac{\dotrho_{{\rm cond}}+\dotrho_{{\rm nucl}}}{\rho}(Y_k-\delta_{k,{\rm cond}}),
\EN
where $\delta_{k,{\rm cond}}$ is unity for the condensing specie, while it is zero for all other $k$'s.

The source due to chemical reactions is given
by\footnote{Note that we have here corrected mistakes with the indices
of $\rho$, $m$, and $\nu$ that were made in the corresponding equation of
Babkovskaia et al.\cite{Babkovskaia2011}.}
\EQ
\dot{\omega}_k=M_k\sum_{s=1}^{N_r}({\nu_{ks}^{''}-{\nu_{ks}^{'}}})
\left[
 k_s^+\prod_{j=1}^{N_s}C_j^{\nu_{js}^{'}}
-k_s^-\prod_{j=1}^{N_s}C_j^{\nu_{js}^{''}}
\right],
\EN
where $C_j$ is the concentration of species $j$, the rate constant of reaction $s$ is given by $k_s=A_sT^{n_s}\exp{(-E_s/RT)}$,
and the values of $A_s$, $n_s$, and $E_s$ are tabulated expressions found in the relevant reaction mechanisms. 
Furthermore, $\nu_{ks}'$ and $\nu_{ks}''$ are the stoichiometric coefficients of the reactant and the product side, respectively.
In most of this work, we will be using the reaction mechanism of 
Panjwani \& Olsen \cite{Panjwani2013}, which is listed in \Tab{tab:panjwani}.
Note that the reactions related to the nitrogen-containing elements have been omitted in the current work. (Another possible sub-set of SiO reactions may be found in Jachimowski and McLain\cite{Jachimowski1983}, although this mechanism also contains a large number of Si-H-O species,
so it is not suitable to extract just a subset of these.)

\begin{table*}[t!]
\begin{center} 
\caption{Reaction mechanism of Panjwani without NOx elements. All third body efficiencies are set to unity.}
\label{tab:panjwani}
    \begin{tabular}{rc|rrr}
    i &Reaction &$A\quad$ &$n\quad$&$E\quad$\\
    \hline
1&$\CO+\OO\Leftrightarrow \COO+\Ox$    &$2.50\times 10^{12}$ &   0.00  &  47800.0\\
2&$\OH+\OH\Leftrightarrow \HHO+\Ox$    &$3.57\times 10^{04}$ &   2.40  & $-2110.0$\\
3&$\CO+\OH\Leftrightarrow \COO+\Hy$    &$4.76\times 10^{07}$ &   1.23  &     70.0\\
4&$\Hy+\OO\Leftrightarrow \Ox+\OH$     &$2.65\times 10^{16}$ & $-0.67$ &  17041.0\\
5&$\Hy+\Ox+\M\Leftrightarrow \OH+\M$   &$5.00\times 10^{17}$ & $-1.00$ &      0.0\\
6&$\Ox+\Ox+\M\Leftrightarrow \OO+\M$   &$1.20\times 10^{17}$ & $-1.00$ &      0.0\\
7&$\CO+\Ox\Leftrightarrow \COO$        &$1.80\times 10^{10}$ &   0.00  &   2385.0\\
8&$\Hy+\OH+\M\Leftrightarrow \HHO+\M$  &$2.20\times 10^{22}$ & $-2.00$ &      0.0\\
9&$\OO+\SiO\Leftrightarrow \Ox+\SiOO$  &$2.31\times 10^{13}$ &   0.00  &  26016.0\\
10&$\OH+\SiO\Leftrightarrow \Hy+\SiOO$ &$1.80\times 10^{10}$ &   0.78  &   1218.0\\
    \end{tabular}
\end{center}
\end{table*}

The equation for fluid temperature $T$ is given by \cite{Babkovskaia2011}
\EQA
\frac{DT}{Dt}&=&\frac{1}{c_v}\left[
\sum_k \frac{DY_k}{Dt}\left(\frac{RT}{M_k}-h_k\right)-\frac{RT}{\bar{M}}\nabla\cdot \UU \right. \nonumber \\
&+&\left. 2\nu \SSSS^2-\frac{\nabla\cdot \qq}{\rho}+\frac{Q_{\rm conv}}{\rho}
\right],
\ENA
where $c_v$ is the heat capacity at constant volume, $R$ is the universal gas constant, $M_k$ is the molar mass of species $k$, 
$\bar{M}$ is the mean molar mass of the gas, $h_k$ is the enthalpy of species $k$, $q=\sum_k h_k \JJ_k - \lambda\nabla T$ is the diffusive heat flux and
$\lambda$ is the thermal diffusivity.
Since the particles we are interested in here are very small, the convective time scale is extremely short, which requires very short time steps.
In order to avoid the problem of too small time steps for the simulation, we make the assumption that particles and fluid are always in thermal equilibrium. This is indeed a good assumption since the convective timescale is much shorter than the time-scales associated with radiative or conductive heat transfer.
Based on this assumption, we now set the particle temperature equal to the fluid temperature while the convective heat transfer from the particles to the fluid is set equal to the sum of the radiative heat transfer and the release of latent heat due to the condensational phase change:
\EQ
Q_{\rm conv}=
\frac{\dotrho_{\rm nucl}\Delta H}{M_{\SiOO}}
+\sum_{j=1}^{N_{\rm part,cell}} 
\left[
q_{{\rm rad},j}+q_{{\rm latent},j}
\right]\nsw_j.
\EN
In this simplified framework, it is implicitly assumed that the total heat capacity of the particles is much lower than the heat capacity of the fluid.
If this assumption does not hold, the thermal inertia of the particles should be incorporated into the temperature equation for the fluid.

\section{Physical parameters}
In this section we will discuss some important physical parameters and try to estimate their sensitivity on the results for the conditions of interest.
\subsection{Surface energy}
There are several papers presenting surface energies of $\SiOO$, but most of the studies are at too low temperatures for the current interest, e.g. \cite{Mizele1985}
Gonzalez-Farina et al. \cite{gonzalez2020} refer to internal communication with Elkem when they use a surface energy of
$\gamma=3.2\times 10^{-2}$~J m$^{-2}$. This does, however, rely on a misinterpretation of measurement data as this is the \textit{dispersive} component of the surface energy measured by inverse gas chromatography \cite{Elkem2000}. 
The dispersive component is often a small fraction of the total surface energy in polar polymers such as silica fume.  
This is confirmed by Overbury et al. \cite{Overbury1975}, who reference Popel et al. \cite{Popel1969} when providing a value of $\gamma_0=0.390$~J m$^{-2}$.
Furthermore, Kingery \cite{Kingery1959} present a temperature-dependent expression for the surface energy, which is consistent with the value of Popel et al.: 
\EQ
\gamma=\gamma_0+\gamma_T(T-T_\gamma), 
\EN
where $\gamma_0=0.307$~J m$^{-2}$, $\gamma_T=3.1\times 10^{-5}$ Jm$^{-2}$K$^{-1}$, and $T_\gamma=2073$ K \cite{Kingery1959}.
\begin{figure}[t!]\begin{center}
\includegraphics[width=\columnwidth]{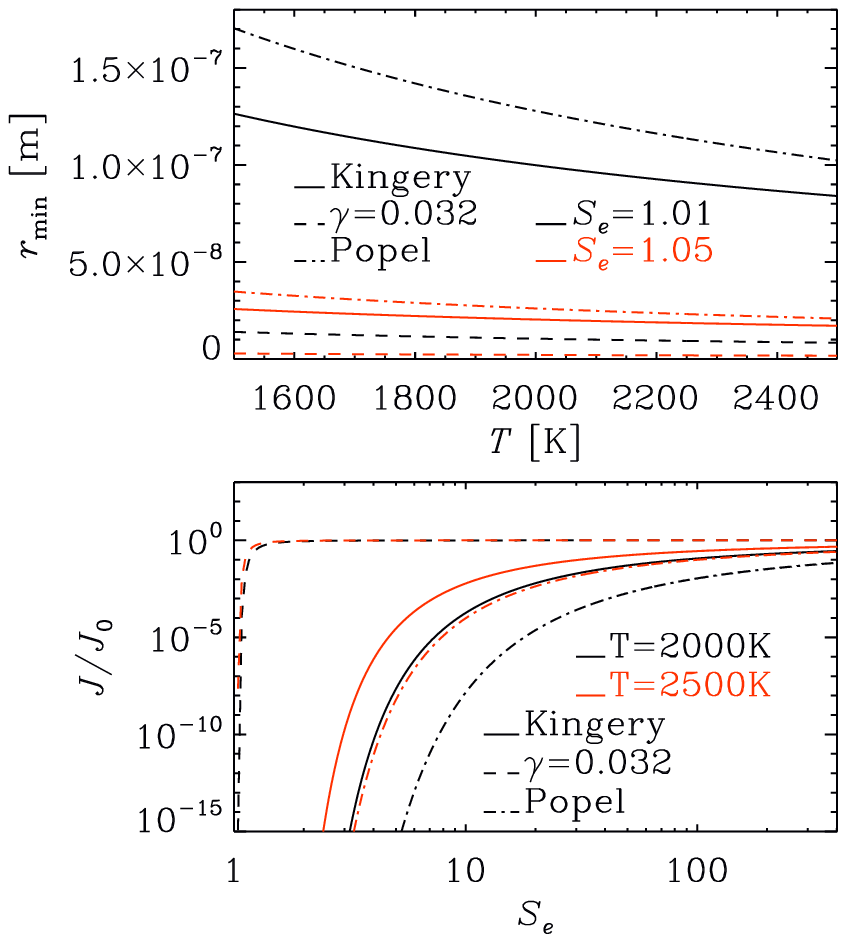}
\end{center}\caption[]{
Upper panel: nucleation radius as a function of temperature.
Lower panel: the exponential part of \Eq{nucleation_rate}.
}\label{gam_surf_energy}\end{figure}
In the upper panel of \Fig{gam_surf_energy}, we see that an increase in surface energy results in an increase in the nucleation radius, but this is only a linear effect. 
The strongest effect of the surface energy is found in the nucleation rate.
In the lower panel of \Fig{gam_surf_energy} we show the value of the exponential term in \Eq{nucleation_rate}, and we see that the effect of the surface energy is enormous -- in particular
for lower super saturation ratios.

\subsection{The nucleation pre-exponential factor}
An important parameter when calculating the nucleation rate from \Eq{nucleation_rate} is the nucleation pre-exponential factor.
In the paper by Gonzalez-Farina et al.\ \cite{gonzalez2020}, the authors set
$J_0=10^{25}$~m$^{-3}$s$^{-1}$ with reference to Lothe \& Pound \cite{Lothe1962}.
It is indeed correct that Lothe \& Pound give the value of  $10^{25}$, but they work in cgs units -- so this value corresponds to $J_0=10^{31}$~m$^{-3}$s$^{-1}$. However, it 
has been known already since the paper of Becker \& Doring \cite{Becker1935,Oxtoby1992} in 1935 
that $J_0$ is variable with respect to vapor concentration:
\EQ
J_0=
\left(
\frac{2\gamma}{\pi m_{\SiOO}}
\right)^{1/2}
v_c (C_{\SiOO}N_A)^2,
\EN
where $N_A$ is Avogadros number. The same pre-exponential factor
is also obtained in the book of Friedlander \cite{Friedlander2000}.
In his review paper, Oxtoby \cite{Oxtoby1992} argues that the nucleation pre-exponential factor of Becker \& Doring should be modified to include an inverse supersaturation factor to read
\EQ
J_0=
\left(
\frac{2\gamma}{\pi m_{\SiOO}}
\right)^{1/2}
\frac{v_c (C_{\SiOO}N_A)^2}{S_e}.
\EN
It should also be mentioned that Zhou et al. \cite{Zhou2014} present an alternative expression for $J_0$, 
\EQ
J_0=
\left(
\frac{2\gamma}{\pi m_{\SiOO}}
\right)^{1/2}
\frac{P_{\rm SiO2}x_{\SiOO}}{k_B T},
\EN
with reference both to the work of Friedlander \cite{Friedlander2000} and the work of Becker \& Doring \cite{Becker1935}, but it is not clear how this expression is obtained from those two references. We therefore omit it from this work. 

\begin{figure}[t!]\begin{center}
\includegraphics[width=\columnwidth]{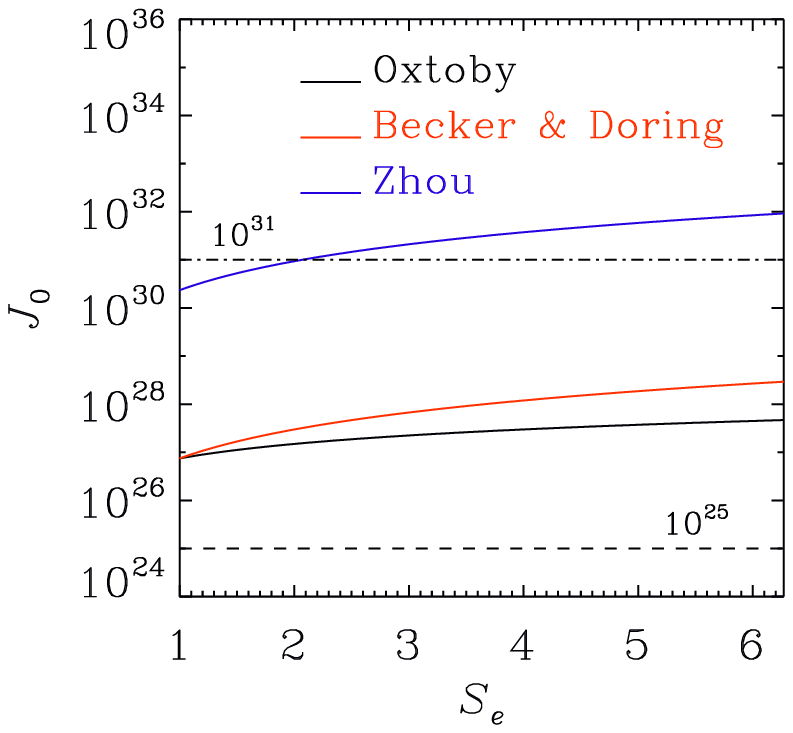}
\end{center}\caption[]{
Nucleation pre-exponential factor, $J_0$, for different models.
}\label{J_S6}\end{figure}

The difference between the above models are shown in \Fig{J_S6}. Based on these results, we will in the following use the model of Oxtoby.
As an interesting side note, for some of the nucleation models (not shown here) the rate of mass nucleation,
$\dot{m}=4\pi r_{\rm min}^3 \rho_{\ms}J/3$, actually goes \emph{down} with increasing $S_e$ for $S_e>2$.

\subsection{Saturation concentration}
\label{conc_sat}
\begin{figure}[t!]\begin{center}
\includegraphics[width=\columnwidth]{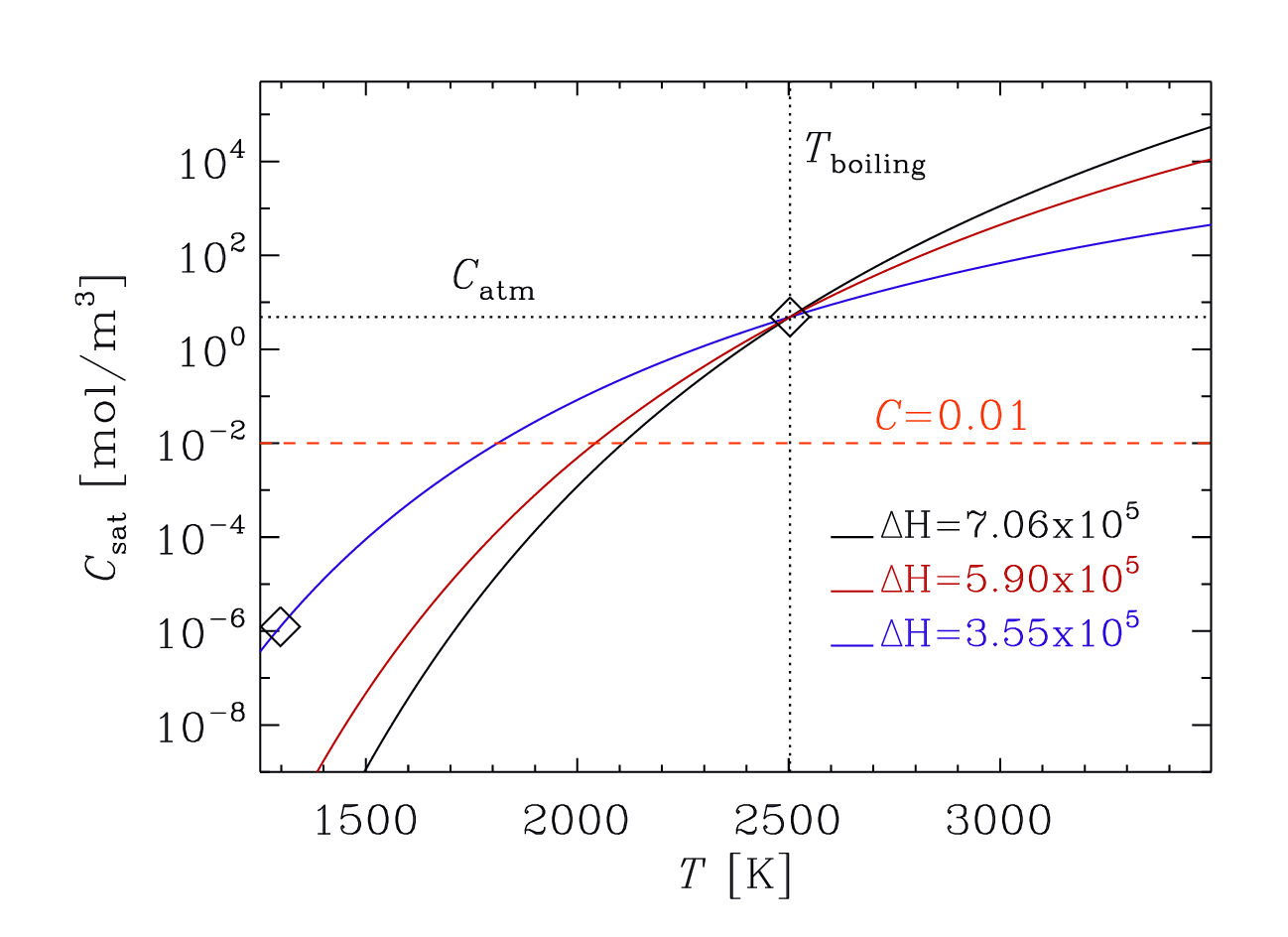}
\end{center}\caption[]{
Here, saturation pressure (upper panel)
and saturation concentration (lower panel) are shown as functions of temperature. The red dashed line correspond to the value 
used in previous papers \cite{gonzalez2020,Vachaparambil2022}.
The two diamonds represent the pressure set-points at (1025 $\degree$C, $10^{-4}$ Torr) and (2503 K, 101325 Pa).
}\label{saturation_concentration}\end{figure}
In the following, the Clausius \& Clapeyron relation will be used to find the saturation concentration of $\SiOO$ as a function of temperature. The saturation pressure is then given by
\EQ
\label{Psat}
P_{\rm sat}=P_{\rm boil} \times \exp \left[ \frac{-\Delta H}{R_{\rm gas}}\left(\frac{1}{T}-\frac{1}{T_{\rm boil}}\right)\right],
\EN
where the subscript ``boil'' refers to the condition at the boiling temperature.
For silica fume at atmospheric pressure, these values are $P_{\rm boil}=101325$Pa and $T_{\rm boil}=2503$K.
The enthalpy of vaporization can be found from different sources to be 
$\Delta H=7.06\times 10^{5}$J/mol \cite{Kraus2012}, $\Delta H=5.99\times 10^{5}$J/mol \cite{Lim2002} or $\Delta H=5.78\times 10^{5}$J/mol \cite{Aylward2002}.

Assuming an ideal gas, the saturation concentration is found from $P=CR_{\rm gas}T$ as
\EQ
\label{Csat}
C_{\rm sat}=\frac{P_{\rm sat}}{R_{\rm gas}T},
\EN
where $R_{\rm gas}=8.314$ J/mol/K is the universal gas constant.
The solid black and dark-red lines in \Fig{saturation_concentration} shows the saturation concentration 
as found from \Eq{Csat} when the enthalpy of vaporization has been set to $\Delta H=7.06\times 10^{5}$J/mol \cite{Kraus2012} and $\Delta H=5.99\times 10^{5}$J/mol, respectively. 
The dotted lines represent the conditions at the boiling point, while the dashed red line corresponds to the value 
used in previous papers \cite{gonzalez2020,Vachaparambil2022}. 

Furthermore, it is reported that the saturation pressure of $\SiOO$ is $10^{-4}$ Torr at 1025 $\degree$C.
Using this reference point in \Eq{Psat} together with the boiling condition yields an enthalpy of vaporization of $\Delta H=3.55\times 10^{5}$J/mol. 
Since 1025 $\degree$C is below the melting point of silica fume, it is, however, very questionable how applicable it is to use this as a reference point.
The resulting saturation pressure and concentration are nevertheless represented by the solid blue lines in \Fig{saturation_concentration}.
For the remainder of this work, we use $\Delta H=7.06\times 10^{5}$J/mol when calculating the saturation pressure.

\section{Validations}
\subsection{Lagrangian vs. Eulerian particle tracking}

\begin{table}[t!]
\begin{center} 
\caption{Lagrangian and Eulerian test simulations. For the eulerian simulations, $n_{\rm bins}$ represents the number of bins that is solved for in the simulation, while for the Lagrangian simulations it represent the number of bins used when calculating the pdf of the resulting particle sizes. Furthermore, $N_{\rm sw}$ is the number of swarm particles that are generated for during the Lagrangian simulations. The total number of physical particles represented by each simulation is given by $N_{\rm phys}$,}
\label{tab:lageul}
    \begin{tabular}{r|rcrcc}
    Case       & $n_{\rm bins}$  &$m_{\rm thr}$         &$N_{\rm sw}$  &$N_{\rm phys}$ & Time\\
               &                 & [kg]                 &              & [$10^{11}$] & [s]\\
    \hline
    Eul. 200   & 200             &    ---               &  ---         & $3.0$       &1.85\\
    Eul. 1200  &1200             &    ---               &  ---         & $3.0$       &8.60\\
    Lagr. 10   &   5             &      $10^{-10}$      &  2           & $2.7$       &0.50\\
    Lagr. 11   &  10             &      $10^{-11}$      &  22          & $3.0$       &0.56\\
    Lagr. 12   &  40             &      $10^{-12}$      &  204         & $3.0$       &1.03\\
    \end{tabular}
\end{center}
\end{table}

\begin{figure}[t!]\begin{center}
\includegraphics[width=\columnwidth]{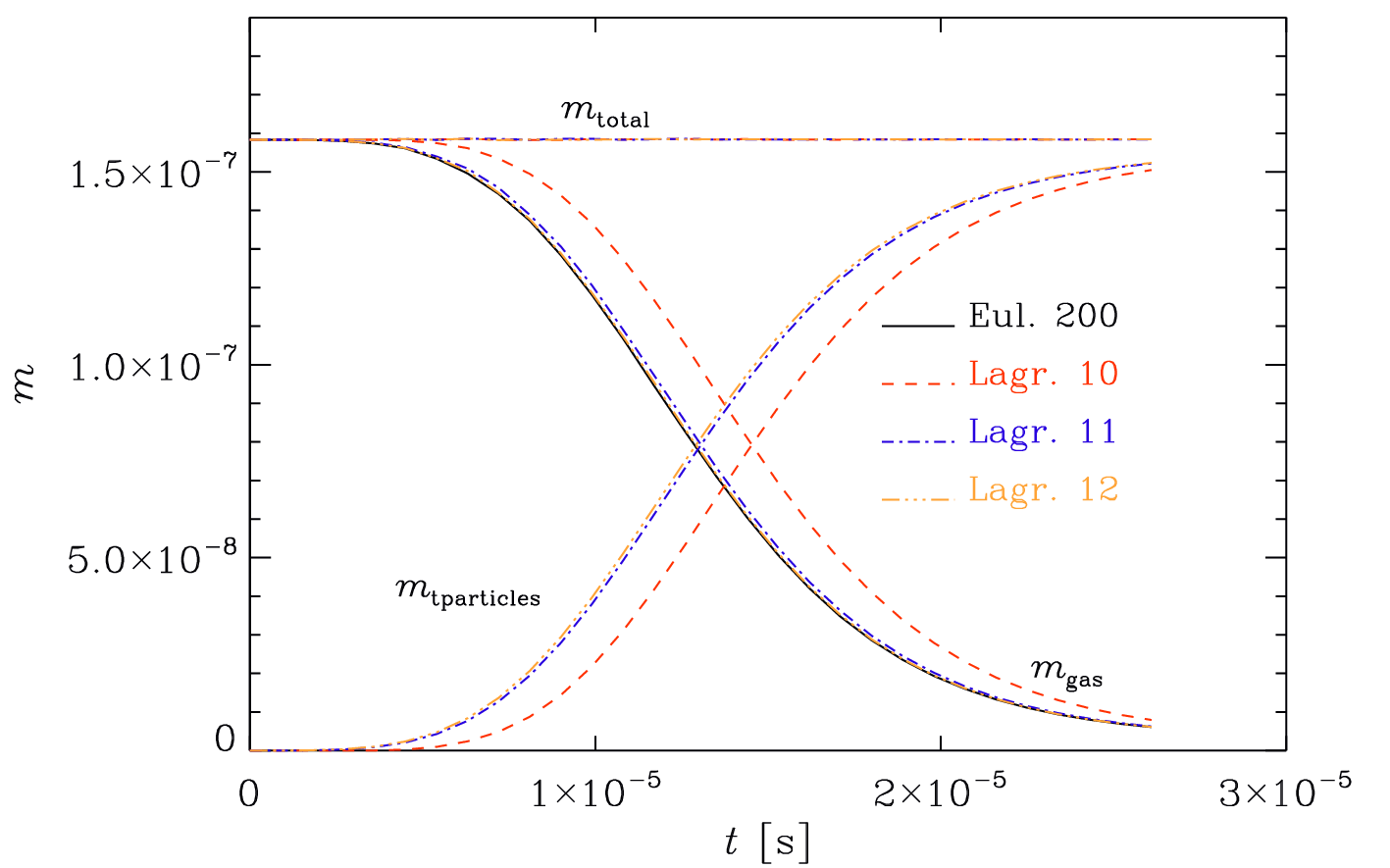}
\end{center}\caption[]{
Mass of $\SiOO$ in gaseous and solid phases as a function of time both for the Eulerian and the Lagrangian approaches. 
}\label{pcomp_mass_nucl_new}\end{figure}

\begin{figure*}[t!]\begin{center}
\includegraphics[width=2\columnwidth]{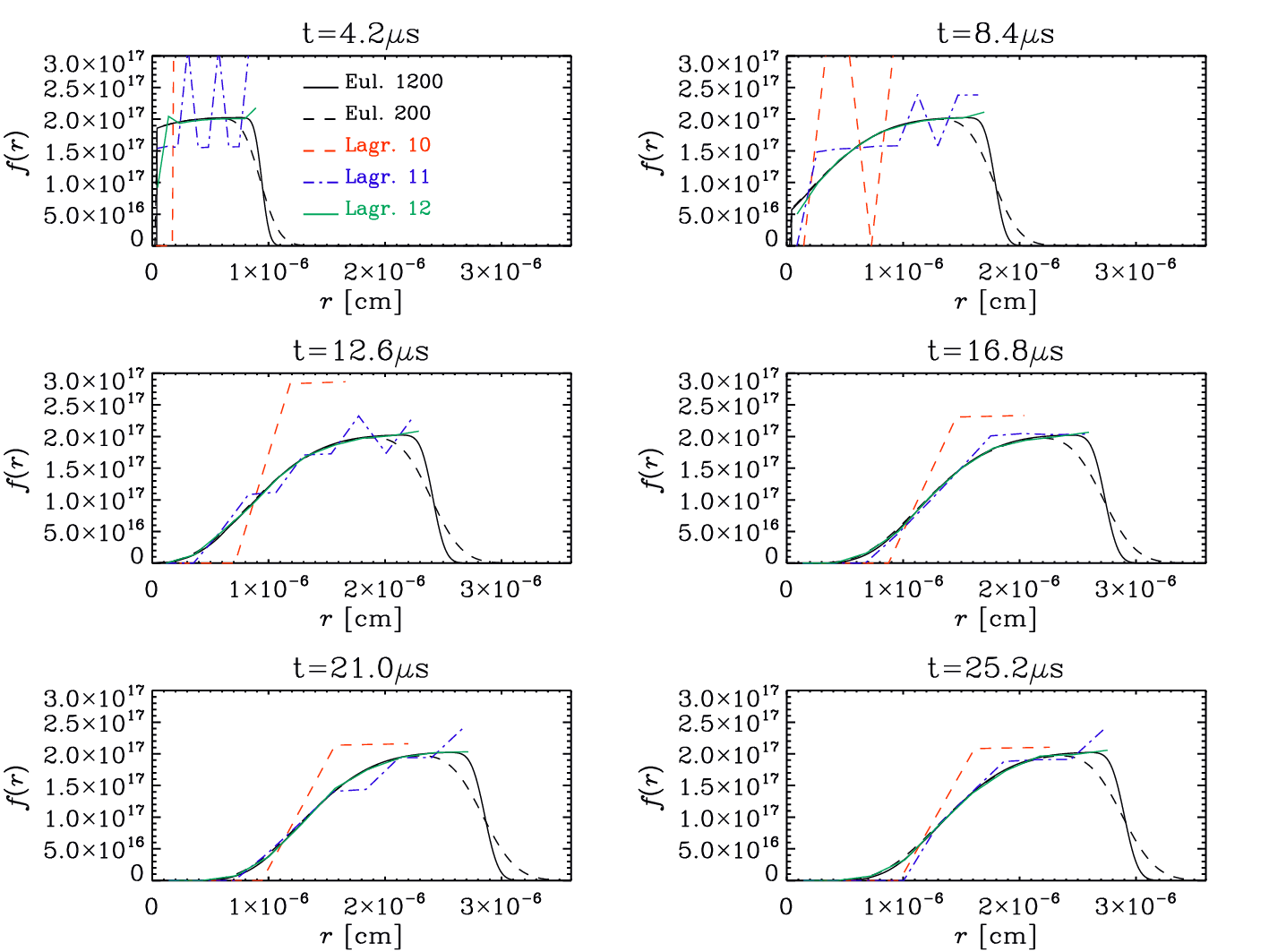}
\end{center}\caption[]{
Particle size distribution for Eulerian and Lagrangian tests.
}\label{pcomp_dist_nucl_new}\end{figure*}

To check the consistency of the two particle approaches (Eulerian and Lagrangian), zero-dimensional nucleation-condensation simulations have been performed for both approaches.
For these non-reactive tests, the initial mass-fractions are $\NN$: 0.767, $\OO$: 0.142 and $\SiOO$: 0.091, and the saturation concentration is set to the same constant value of 0.01 mol/m$^3$ as in Gonzalez-Farina et al.\cite{gonzalez2020}.
In \Fig{pcomp_mass_nucl_new} the evolution for the mass of silicon in the form of particles and gas are shown both for the Eulerian and Lagrangian particle approaches.
For the Lagrangian cases, three different values of $\rho y_{\rm nucl,thresh}$ are tested, and we see that for the current test,
using $\rho y_{\rm nucl,thresh}=10^{-11}$ gives somewhat too slow nucleation and condensation, while decreasing $\rho y_{\rm nucl,thresh}$ below $10^{-11}$ does not improve the results further.
With  $\rho y_{\rm nucl,thresh} \le 10^{-11}$, the Eulerian results are also reproduced.

To get a better understanding of the physics, \Fig{pcomp_dist_nucl_new} shows
the particle size distribution for the Eulerian and Lagrangian
cases. The various panels represent
different times. As we can see, the Lagrangian case with 
$\rho y_{\rm nucl,thresh}=10^{-11}$ or smaller resembles the Eulerian results pretty nicely, although the distribution is rather spiky. 
The case with $\rho y_{\rm nucl,thresh}=10^{-10}$ also has some resemblance with the two other results,
although the fact that this simulation resulted in only two swarm particles means that it is hard to claim that we have a real particle size distribution.
For the latter case, only 5 size bins were used when producing the probability distribution function (pdf), while for the two other Lagrangian cases a total of 10 and 40 size bins were used.
This is nevertheless in stark contrast to the Eulerian case, where 200 and 1200 size bins were used. It is also clear that even with 200 Eulerian size bins, the distribution is not reproducing the larger size particles correctly. From the upper left panel (earliest time), we see that there is an equilibrium between nucleation and convection -- meaning that nuclei that are 
produced are convected upward to larger size bins at the same rate as they are produced.
This results in a flat size distribution that extends towards the right.
Then, as the amount of $\SiOO$ in the gas is reduced, the rate of convection is also reduced, but the nucleation rate is reduced even more.
This results in a relative reduction of particles at small radii, while the distribution is traveling further to the right.

From \Tab{tab:lageul} it can be seen that the CPU-time required for even the smallest Eulerian simulation is longer than for the largest Lagrangian simulation.
In real three-dimensional simulations, the benefit of the Lagrangian approach becomes even more convincing
since an Eulerian simulation will have to solve for all particle sizes even in parts of the domain where there are no particles,
while a Lagrangian simulation use resources on particles only where there really are particles.
For most three-dimensional simulations this may have a large impact. For this reason,
and due to the more accurate particle size distribution and the higher flexibility when it comes to particle content,
the Lagrangian particle approach is considered superior to the Eulerian approach for the work of interest here.

\subsection{Chemical reactions}

\begin{figure}[t!]\begin{center}
\includegraphics[width=\columnwidth]{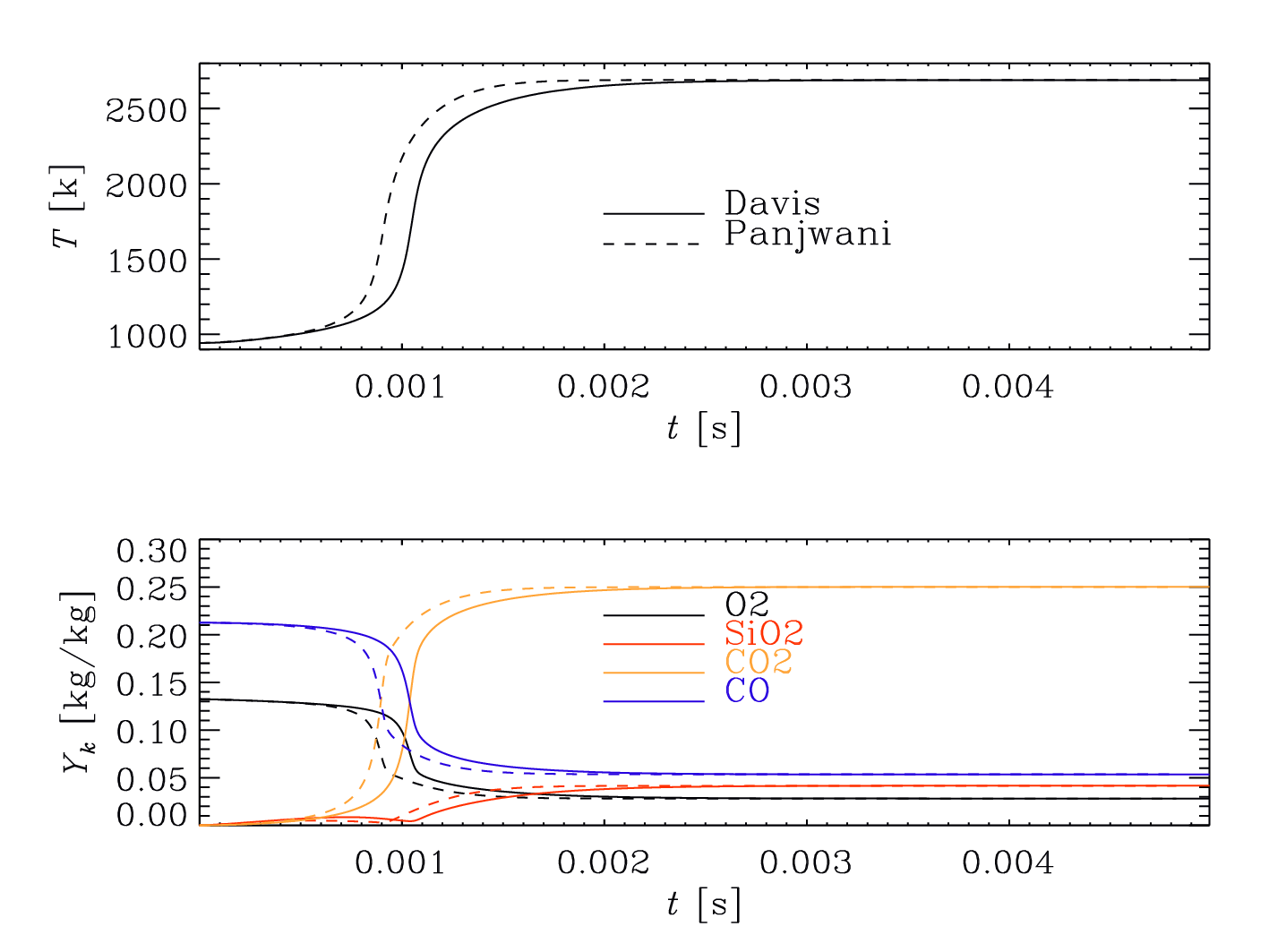}
\end{center}\caption[]{
Comparison of zero-dimensional reactive simulations  obtained with the reaction mechanism listed in \Tab{tab:panjwani} with results obtained with the reaction mechanism of Davis \cite{Davis2005}. Here, reactions 9 and 10 of the Panjwani mechanism have been added to the Davis mechanism. 
}\label{panjwani_Davis}\end{figure}

To validate the non-Si containing reactions of the mechanism presented in \Tab{tab:panjwani},
we perform a zero-dimensional simulation with all the reactions listed in \Tab{tab:panjwani}
and compare the results with those obtained using the Davis \cite{Davis2005} reaction mechanism,
where we have added reactions 9 and 10 listed in \Tab{tab:panjwani} to account for Si reactions.
From the results shown in \Fig{panjwani_Davis}, we see that there are only smaller differences in ignition delay time between the two mechanisms.
This supports the accuracy of the non-Si part of the Panjwani mechanism, and we will therefore continue by using it for the rest of the simulations presented in this document.

\section{Results}

\subsection{Zero-dimensional simulations}
In order to be able to emphasize the more fundamental aspects of silica
fume formation, zero-dimensional simulations are performed here, which
means that there are no spatial gradients.
This corresponds to premixed conditions.

\begin{figure*}[t!]\begin{center}
\includegraphics[width=2\columnwidth]{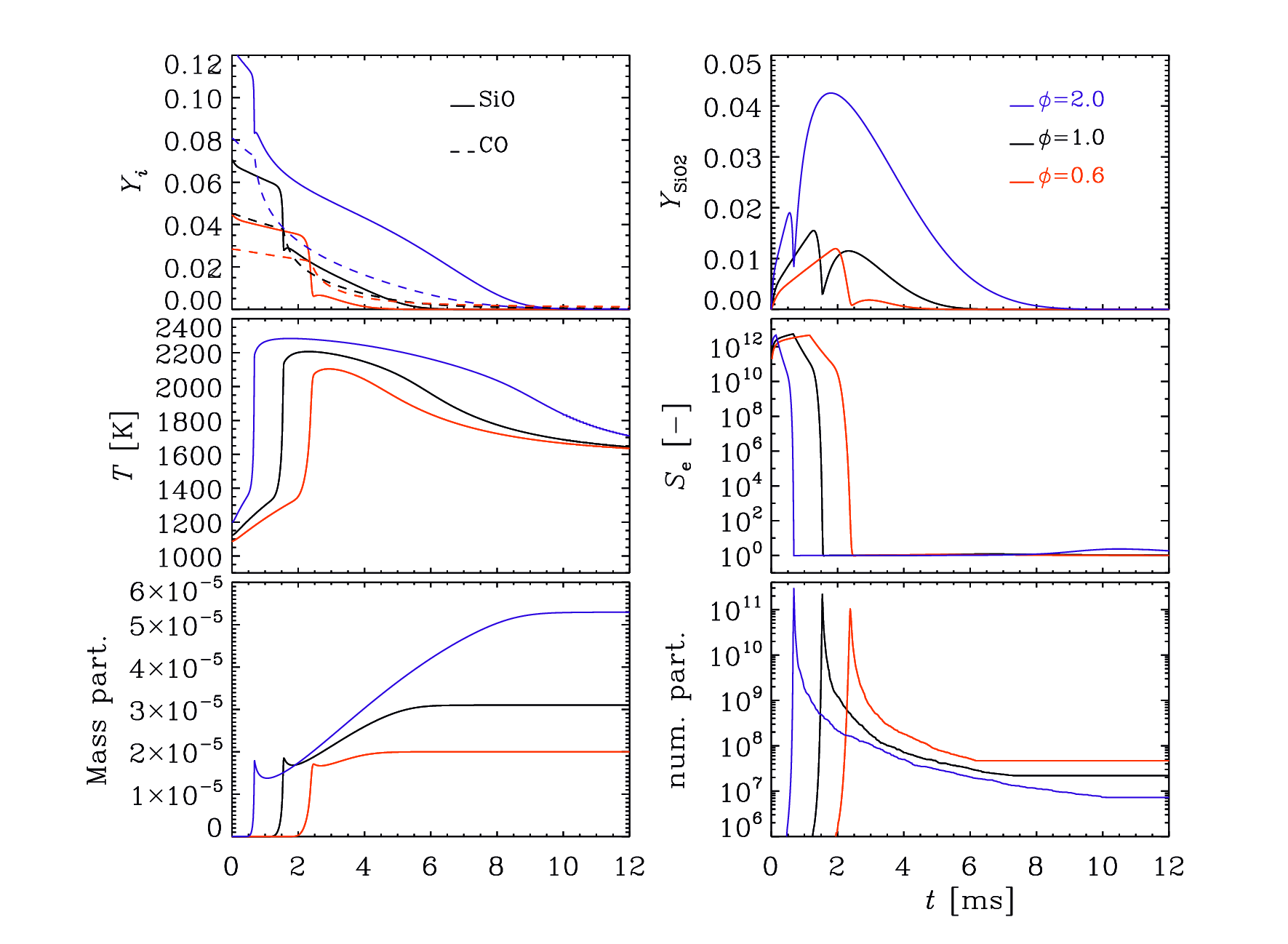}
\end{center}\caption[]{
Time evolution for various variables for equivalence ratios of 1.0 (black), 0.6 (red) and 0.3 (blue).
}\label{part_evo_coag}\end{figure*}

\begin{figure*}[t!]\begin{center}
\includegraphics[width=2\columnwidth]{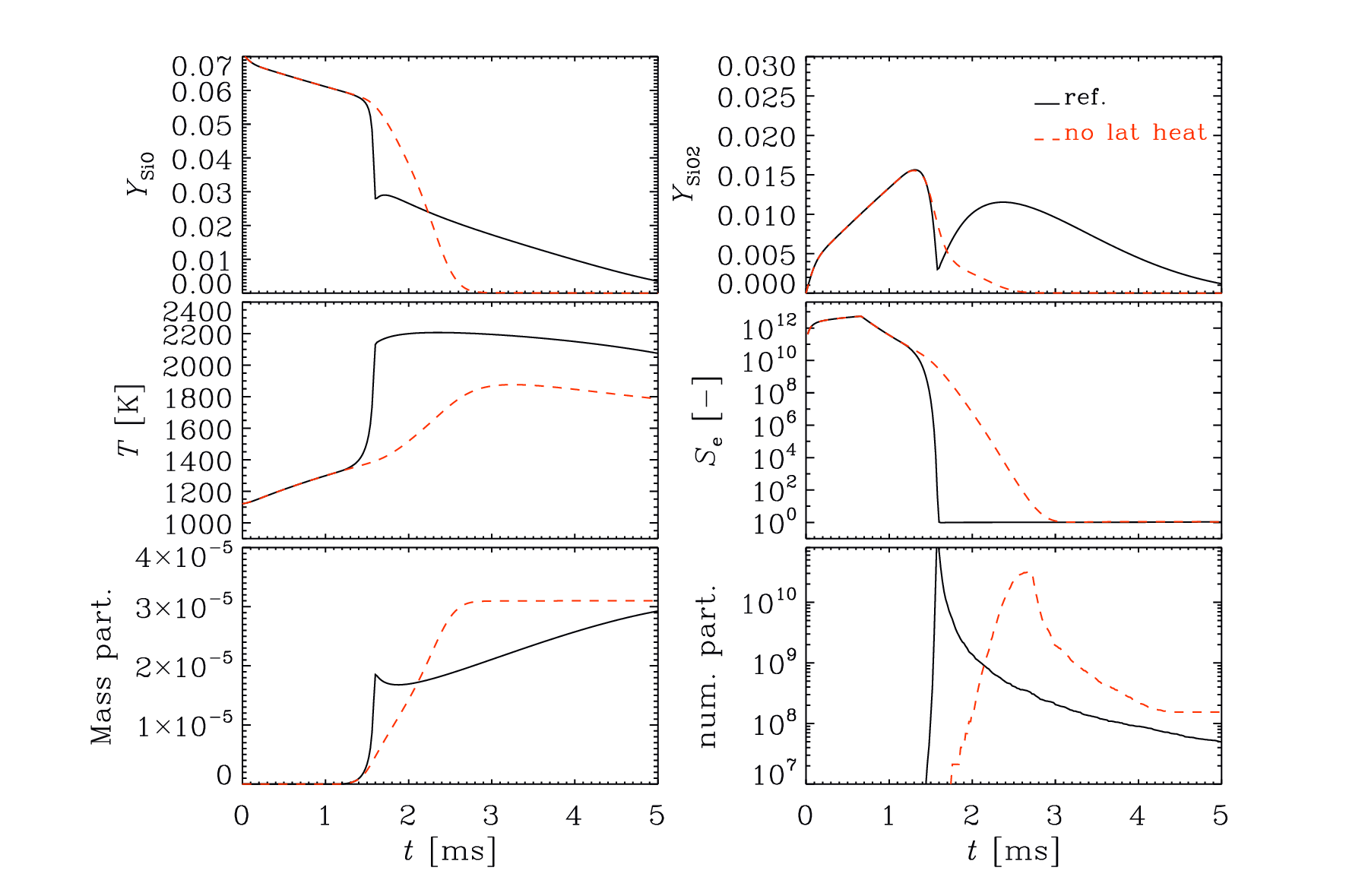}
\end{center}\caption[]{
Time evolution for various variables for equivalence ratios of 1.0 with and without liberation of latent heat during condensation and nucleation.
}\label{part_evo_coag_latentheat}\end{figure*}

Under real multi-dimensional conditions, entrainment of cold air into the
combustion products will cool down the product stream.
Therefore, to emulate this effect in the zero-dimensional simulations,
an explicit cooling term is introduced. This cooling term is given by
\EQ
\label{eq:cool}
Q_{\rm cool}=\sigma_{\rm SB}a_{\rm op}(T_{\rm surr}^4-T^4),
\EN
where $T_{\rm surr}=1600$ K and $a_{\rm op}=1$ for the reference case.

\subsection{Growth of silica fume particles}
In \Fig{part_evo_coag}, the evolution of six different variables is presented as a function of time for three different equivalence ratios, where the equivalence ratio is defined as
the actual fuel to air ratio divided by the stoichiometric fuel to air ratio, i.e.,
\EQ
\phi=\frac{F/A}{(F/A)_{\rm stoich}}.
\EN
Let us first consider the results for stoichiometric conditions ($\phi=1$) represented by the black lines, which from now on is referred to as the reference case. 
From the upper left panel, an initial, almost constant, decrease of the mass fractions of $\SiO$ and $\CO$ can be seen.
This lasts until around 1.3 ms when a sudden drop in $\SiO$ is observed. 
By inspecting the other panels of the same figure, it is clear that a similar sudden change occurs also for the other variables at the same time. 
From the two last panels it is clear that this abrupt change coincides with the appearance of the first particles,
which for the reference case happens at around 1.25 ms. This is also the time when the condensation rate of gaseous $\SiOO$ on the particles becomes non-zero (not shown). 

In this context, the condensation has two main effects, the first is the liberation of significant amounts of thermal energy due to the phase change from gaseous to liquid state, which results in a temperature increase. The second effect is a shift in the equilibrium of the $\SiO$ oxidation reactions (reactions 9 and 10 in \Tab{tab:panjwani}), which results in increased oxidation rates and therefore also increased temperature. The shift in the equilibrium of the $\SiO$ oxidation reactions is due to the transfer of $\SiOO$ from the gas phase to the particles.
From \Eqs{drpdt}{nucleation_rate} it is clear that higher temperature results in higher nucleation rate and faster condensation. This means more particles for gaseous $\SiOO$ to condense on and faster condensation on each particle, which again results in an even faster temperature increase due to the release of the enthalpy of evaporation and increased oxidation of \SiO. 
The result of this is a runaway chain reaction, which does not stop until the amount of gaseous $\SiOO$ has been reduced so much that it equals the equilibrium concentration at that temperature.
This corresponds to a supersaturation ratio of $\SiOO$ ($S_e$) at or below unity.
As can be seen from the right panel of the second row of \Fig{part_evo_coag}, this occurs at around 2.5~ms for the reference case ($\phi=1$).

To elaborate a bit more on this, the effect of the latent heat becomes clear by considering \Fig{part_evo_coag_latentheat},
where the dashed red line represents the exact same simulation as the reference case (black solid line), except that the latent heat is set to zero.
Here we see that the temperature increase is significantly slower when the latent heat is zero than for the reference case.
The changes in the other parameters visualized in the figure are also less steep, but they are nevertheless far from flat.
This remaining steepness comes from the increased oxidation rate of $\SiO$ due to the change in the equilibrium of the $\SiO$ oxidation reactions when $\SiOO$ condense on the particles.

\begin{figure}[t!]\begin{center}
\includegraphics[width=\columnwidth]{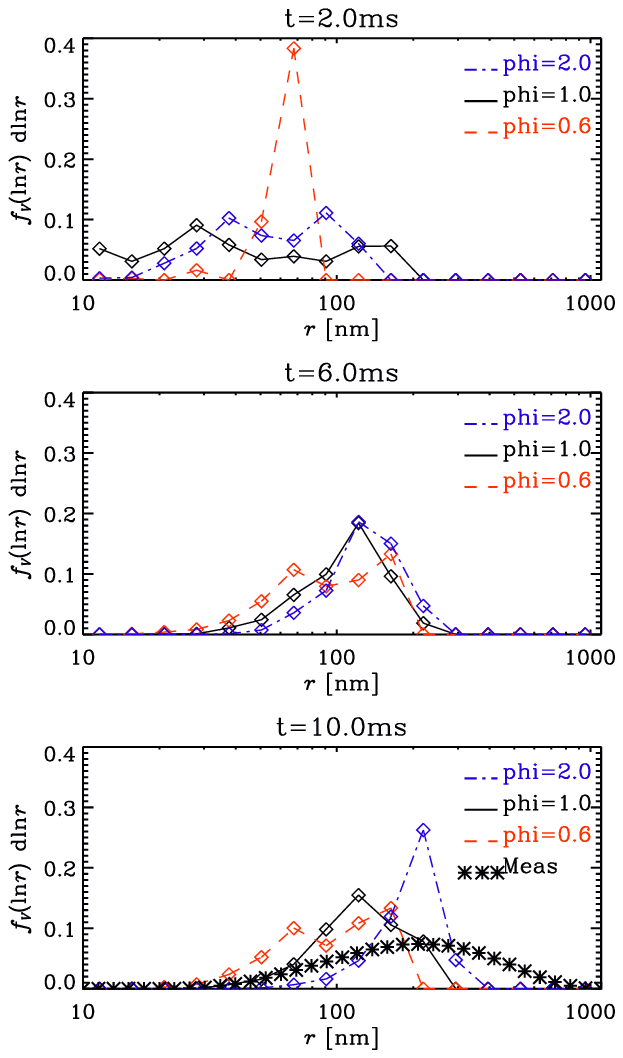}
\end{center}\caption[]{
Particle size distributions for the same simulations as presented in \Fig{part_evo_coag}. The symbols in the lower right panel corresponds to measured particle size distributions.
}\label{pcomp_dist_coag}\end{figure}

\begin{figure}[t!]\begin{center}
\includegraphics[width=\columnwidth]{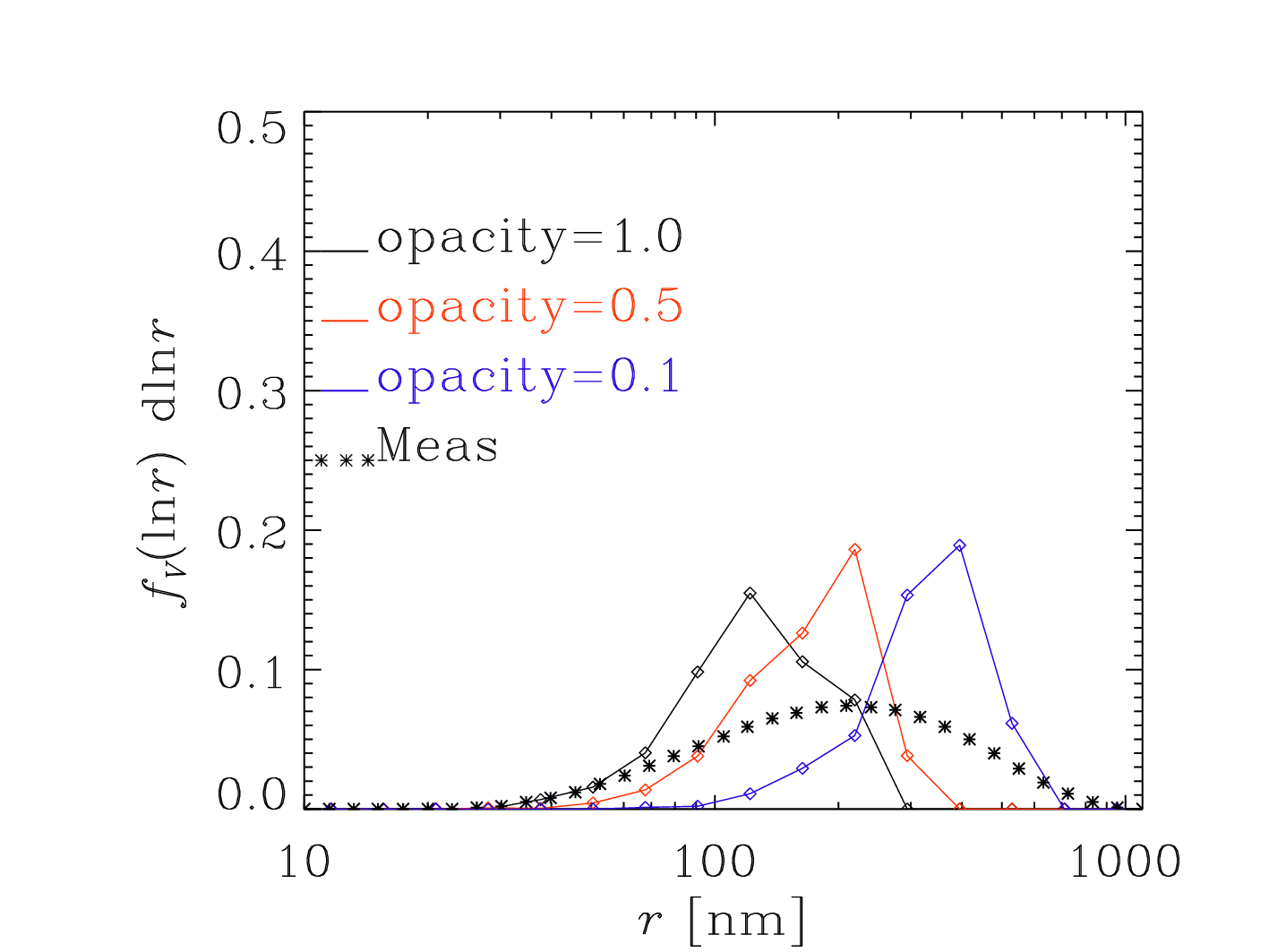}
\end{center}\caption[]{
Final (after temperature has decreased below the melting temperature) particle size distributions for simulations with different cooling rates ($a_{\rm op}=$ 1 (black), 0.5 (red), 0.1 (blue)). The black stars represent the same measurements as presented in \Fig{pcomp_dist_coag}.
}\label{pcomp_dist_coag_cool}\end{figure}

Looking back at \Fig{part_evo_coag}, the amount of $\SiO$ continues to gradually decrease after the previously discussed chain-reaction has stopped.
This yields an increase in particle mass due to condensation of gaseous $\SiOO$. At the same time, the lower right panel shows that the total number of particles is drastically decreasing while the total particle mass increases. The reason for this is particle coagulation due to Brownian motion of the particles. This coagulation, with its corresponding reduction in particle number and increase in average particle diameter, continues until the temperature has decreased down to the melting temperature of silica fume at 1560 $\degree$C (=1833 K), which happens after about 7.2 ms for the reference case.

By considering the results for other equivalence ratios in \Fig{part_evo_coag} (red and blue lines), it is clear that they follow the same trend as the stoichiometric case. One notable difference is the fact that the lowest equivalence ratio, while it ends up with the lowest total particle mass, still has the highest number of particles. This is due to the fact that, because of its lower amount of chemical energy, it is cooled down to the melting temperature of silica fume before coagulation has had time to grow the particles to the same size as for the larger equivalence ratios.

\subsection{Particle size distributions}

It is interesting to look at the particle size distribution for different times for the same simulations as presented in \Fig{part_evo_coag}.
For the simulation with $\phi=0.6$, only the very first small nuclei have started to appear at $t=2.0$~ms.
Since these nuclei are generated in the presence of an extremely high super-saturation ratio, they will grow very rapidly.
This results in the narrow peak for the case with $\phi=0.6$ in the upper panel of \Fig{pcomp_dist_coag}. While for the two higher equivalence ratios, the particle nucleation and growth process has already been going on for a while at $t=2$ ms, which means that the supersaturation ratio has decreased to around unity.
Hence, the particles present in these to simulations at this time have different age and have experienced very different conditions during their life time,
resulting in much broader particle size distributions for the two larger equivalence ratios at $t=2$~ms.
Going to the last panel in the figure, for $t=10$ ms, it is clear that the particle size distributions are shifted more and more to larger particles as the equivalence ratio is increased. The reason for this can be found from \Fig{part_evo_coag}, where we see that for larger equivalence ratios, the simulation stays longer at temperatures above the melting temperature of silica fume. This results in more time for particles to coalesce due to Brownian motion, and therefore larger particles. In the lower panel of \Fig{pcomp_dist_coag}, data measured in a real furnace is also presented. Although the measured distribution is broader than the simulated ones, and shifted towards larger particles, the difference is not dramatic given the zero-dimensional premixed approximation made in the simulations presented here.

The final particle size distribution depends on the time that the particles stay above the melting temperature. The longer the particles are above this temperature, the more coagulation will occur, leading to a shift of the particle size distribution towards larger particles. This is visualized in \Fig{pcomp_dist_coag_cool}, where the final particle size distributions of three simulations with different cooling rates are compared. Large opacity ($a_{\rm op}$ in \Eq{eq:cool}) means high cooling rate. 
The particle size distributions obtained from the numerical simulations (solid lines) are all narrower than the measured particle size distribution (symbols). The reason for this is that all the particles grown in a zero-dimensional premixed numerical simulation will experience the same equivalence ratio, while in reality the local equivalence ratio where the different particles grow will vary within the domain. This variation will result in a broader particles size distribution, as can be understood by inspecting the results presented for different equivalence ratios in \Fig{pcomp_dist_coag}. The particle size distribution was measured using a Malvern Panalytical Mastersizer 3000 laser diffraction particle size analyzer. The sample was obtained from the bag filter at one of Elkem’s silicon furnaces.

\begin{figure*}[h!]\begin{center}
\includegraphics[width=1.9\columnwidth]{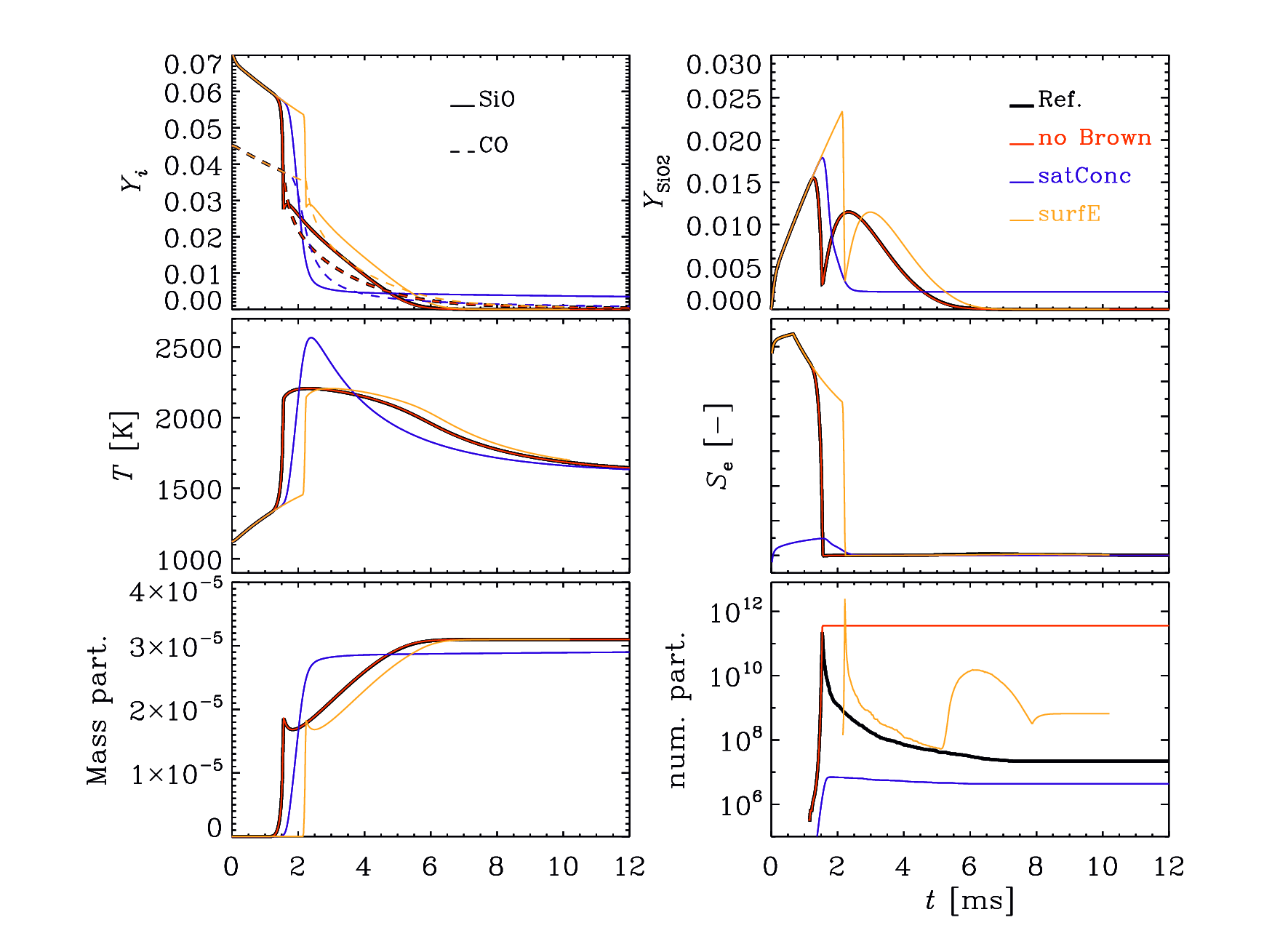}
\end{center}\caption[]{
Black line represent the reference case with equivalence ratio of unity, as also presented in \Fig{part_evo_coag}. The effect of turning off Brownian motion in the reference case is shown from the red line, while if the saturation concentration in the reference case is set to a constant value of $10^{-2}$ mol/m$^3$, the blue line is obtained. Finally, reducing the surface energy of the reference case down to $\gamma=3.2\times 10^{-2}$~J m$^{-2}$ yields the yellow line.  
}\label{part_evo_gamma}\end{figure*}

\subsection{Sensitivity to saturation concentration, surface energy, and Brownian motion}

It is interesting to investigate the sensitivity of the final results to some of the more important physical parameters.
Here we will consider the following parameters: saturation concentration of $\SiOO$, surface energy of silica fume, and Brownian motion. 

\begin{figure}[h!]\begin{center}
\includegraphics[width=\columnwidth]{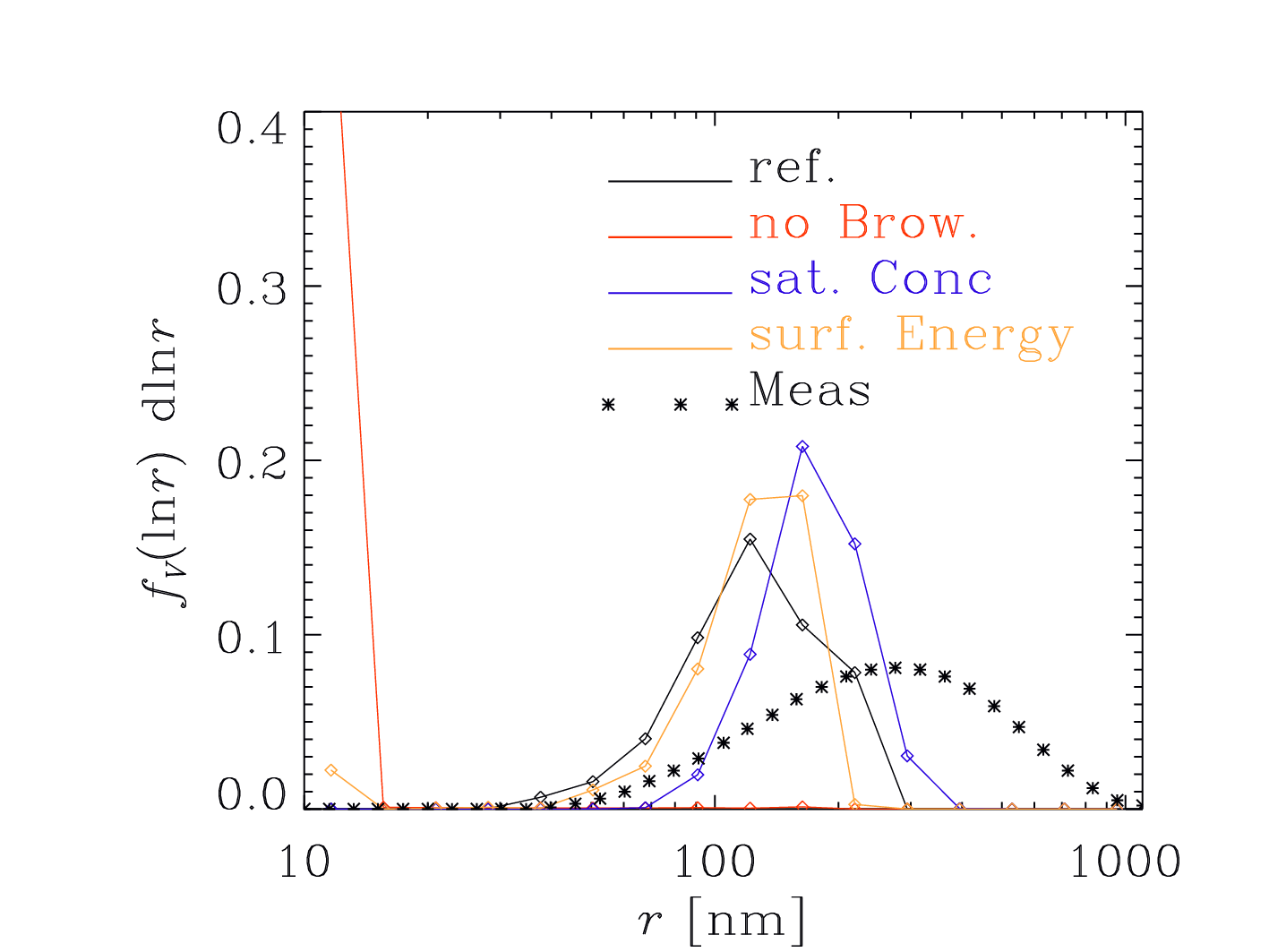}
\end{center}\caption[]{
Final particle size distributions for the same simulations as presented in \Fig{part_evo_gamma}.
}\label{pcomp_dist_coag_div}\end{figure}

By setting the saturation concentration of $\SiOO$ to a constant value of 0.01 mol/m$^3$, as done in, e.g., \citet{gonzalez2020} and \citet{Vachaparambil2022},
the results represented by the blue line in \Fig{part_evo_gamma} are obtained. 
By comparing with the reference case (thick black lines), it is clear that the temporal evolutions are strongly dependent on the saturation concentration, even though the final results are not extremely different for this particular setup. 
The most prominent difference is the fact that the supersaturation ratio ($S_e$) is never very large (the peak is roughly ten orders of magnitude smaller than for the reference case), which results in the formation of larger nuclei; see \Eq{rmin}.
It should also be noted that the particle nucleation starts slightly later than for the reference case, but the transfer of $\SiOO$ from the gas phase to the particles proceeds more or less continuously until all Si is in the particles. Finally, we note by considering \Fig{pcomp_dist_coag_div} that the particle size distribution is slightly larger than for the reference case. For this case, coalescence due to Brownian motion is much less important than for the reference case. This is due to the fact that the initial nuclei are much larger when the supersaturation is small, as can be seen from \Eq{rmin}.

Let us now see what happens if Brownian motion is neglected in the reference simulation.
This is represented by the red lines in \Fig{part_evo_gamma}.
For all but the very last panel, the results without Brownian motion follow the reference case very accurately.
The only panel where a difference is found is the one for the number of particles,
where the final number of particles without Brownian motion is found to be a factor of $10^4$ larger than for the reference case.
This is clearly due to the lack of any coalescence for the case without Brownian motion.
This is shown even more clearly in \Fig{pcomp_dist_coag_div}, where it is seen that even at the final time only the smallest particles exist for the simulation without Brownian motion.
This means, as has already been mentioned previously, that the presence of Brownian motion is instrumental in determining the final particle size distribution.
Another interesting aspect with this case is that none of the other parameters presented in the figure seem to be influenced by the particle size.
This is maybe surprising since we know from \Eq{drpdt} that the change in particle size only depends on the temperature and the difference
between concentration of $\SiOO$ and its saturation concentration.
Based on this, one would think that, due to their larger combined surface area, smaller particles should yield faster transfer to the particle phase.
The reason that there is still no effect on the other parameters is that the first part of the condensation ($1.25<t<1.5$ ms) is so fast that essentially no coalescence has time to occur.
For the next phase of condensation  ($t>2$ ms), the supersaturation is almost unity, so all condensation is controlled by how fast $\SiO$ is oxidized to produce $\SiOO$. Even at the lower particle surface area of the reference case, the condensation rate is fast enough to maintain a supersaturation ratio close to unity, meaning that the increased total surface are for the case without Brownian motion has no effect on the condensation rate.

We have already mentioned that the nucleation rate is very sensitive to the surface energy of silica fume.
So, how does a reduction in of the surface energy by a factor of ten influence the final results?
This is studied by setting $\gamma=3.2\times 10^{-2}$~J m$^{-2}$, and the results are presented as the orange curves in \Figs{part_evo_gamma}{pcomp_dist_coag_div}.
From \Eq{nucleation_rate} it seems that smaller surface energy will yield a much larger exponential term of the nucleation rate. Although this is true for lower super-saturation ratios, the exponential term is always about unity for the large super-saturation ratios experienced in the beginning of these simulation. This means that the mass flux of $\SiOO$ from the gas phase to the particles is determined only by the pre-exponential factor of \Eq{nucleation_rate}, which scales as the square root of the surface energy, and the cube of the critical radius; see \Eq{rmin}.
Since the critical radius scales linearly with the surface energy, the mass flux due to nucleation scales with the surface energy to the 3.5 power.
This is why we see that the first particles appear later for the orange line in \Fig{part_evo_gamma}.
The most prominent feature of the orange line is nevertheless the sudden increase in the number of particles appearing at around 5 ms. This increase can be understood by the fact that the super-saturation ratio is now close to unity.
When the fluid temperature starts to decrease because of the cooling, this results in a small increase in the super-saturation ratio,
which results in a significant value of the exponential term of \Eq{nucleation_rate}.
Because of the smaller value of the surface energy, this kick starts nucleation again. 

\section{Conclusions}

Our main conclusions can be summarized as follows:
\begin{enumerate}
    \item For the first time, simulations of silica fume using parameters that makes physical sense have been performed. 
    \item Eulerian and Lagrangian particle tracking methods have been compared. They are found to agree with each other, but the Lagrangian approach is both cheaper and more flexible.
    \item When particle nucleation commences, a temperature-controlled chain reaction is initiated, where the liberation of vapor energy due to nucleation and condensation on particles leads to an increased temperature. Increased temperature speeds up both nucleation and condensation, which again results in an even faster temperature increase. On top of this, the transfer of $\SiOO$ from the gas phase to the particles shifts the $\SiO$-oxidation reactions to producing more $\SiOO$, which again results in an enhanced temperature increase due to the exothermicity of these oxidation reactions.
    \item Coalescence due to Brownian motion governs the overall particle size distributions. Resulting in an average particle size that is two orders of magnitude larger than without this effect. The final particle size distribution is dependent on how long the particles experience temperatures above their melting temperature. Below the melting temperature, coalescence will not occur.
    \item The simulations suggest that CO has the same reactivity as SiO. It should also be mentioned that in the work presented in this paper, there was no water vapor in the inlet gas mixture, and it is known that introducing steam increases the reactivity of CO compared to SiO even further. These aspects motivate the development of more tailored reaction schemes for SiO oxidation.
\end{enumerate}

\newpage
\appendix

\section{Eulerian particle approach}
\label{eulerian_part}

In the Eulerian particle approach, one solves for the distribution of particle sizes in every grid cell.
This distribution can be based on particle radius, $f(r)$, the logarithm of particle mass, $\tilde{f}(\ln m)$, or the logarithm of particle radius, $\tilde{\tilde{f}}(\ln r)$.
Each of these distributions are made up of a certain number of particle bins, where an evolution equation is solved for each of those bins.

With the above definitions of the various particle size distributions, we know that the total particle number density is given by
\EQ
n=\int f dr=\int \tilde{f}d\ln m=\int \dbtilde{f}d\ln r
\EN
such that
\EQ
f dr=\tilde{f}d\ln m=\dbtilde{f}d\ln r
\EN
and
\EQ
f=\frac{3\tilde{f}}{r}=\frac{\dbtilde{f}}{r}.
\EN

Since $\ln r_n=\ln r_0 + n\;d\ln r$ the $n$'th radius in the logarithmic radius case is given by
\EQ
r_n=r_0\exp (n\; d\ln r),
\EN
which means that a nucleus with radius $r_{\rm nucl}$ belongs to bin number
\EQ
n=\int(r_{\rm nucl}-r_0)/dr
\EN
for the linear radius case and
\EQ
n=\int
\frac{\ln(r_{\rm nucl}/r_0)}{d\ln r}
\EN
for the logarithmic radius case. 

In the following, we will use the radius binning, i.e., $f(r)$.
The evolution equation for particle number density is then given by
\EQ
\frac{\partial f}{\partial t}+\nabla\cdot(\vvv_p f)=D_p\nabla^2f-
\frac{\partial}{\partial r}\left(Gf\right)+T_{\rm coll},  
\EN
where $T_{\rm coll}$ is representing particle collisions,
the size dependent particle number density is given by $f=f(t,\xx,r)$,
and the total particle number density is the
integral of $f$ over $r$:
\EQ
n(t,\xx)=\int_0^\infty f(t,\xx,r) dr.
\EN

The sink due to condensation is given by
\EQ
\dotrho_{\rm cond}=\sum_{i=1}^{N_{\rm radius bins}} 4\pi r_i^2 G\rho_{\ms}f_i \delta r
\EN
for the Eulerian approach, where $f_{{\rm cond},j}=4\pi r_j^2 G \rho_{\ms}$.

\subsection{Moments}
The $n$'th order moment is given by
\EQ
m_n=\sum r^n f dr,
\EN
such that $m_0=n$ gives the particle number density, while 
\EQ
m_3=\frac{3}{4\pi V_{\rm domain}}V_{\rm parts}, 
\EN
is a measure the total volume of all the particles residing within a volume of the simulation domain given by $V_{\rm domain}$. 

Mass conservation demands that the total mass of all $\SiOO$ molecules equals the mass of $\SiOO$ molecules in the gas phase plus the mass of those in the particle phase, i.e.,
\EQ
m_{\rm sio2,total}=m_{\rm sio2,gas}+m_{\ms},
\EN
where the mass of $\SiOO$ in the gas phase is given by
\EQ
m_{\rm sio2,gas}=\rho_g Y_{\SiOO} V_{\rm domain}
\EN
and the mass of $\SiOO$ in all the silica fume particles is given by
\EQ
m_{\ms}=V_{\rm parts} \rho_{\ms} = \frac{4}{3}\pi\rho_{\ms}m_3 V_{\rm domain}.
\EN
Here, we have used the fact that the true density of silica fume particles is $\rho_{\ms}=2.196$ g/cm$^3$.

\section{Brownian motion}
\label{app:brown}

\subsection{Derivation of expression for Brownian motion}

For a particle whose radius is smaller than the mean free path of the carrier
gas, its velocity $\mathbf v(t)$ follows the Langevin (Ornstein--Uhlenbeck)
equation
\begin{equation}
  m_p\,\frac{d\mathbf v}{dt}
      = -\frac{m_p}{\tau_p}\,\mathbf v
        + \mathbf F_B(t),
  \label{eq:langevin}
\end{equation}
where $\tau_p$ is the momentum‐relaxation time (Epstein drag) and
$\mathbf F_B(t)$ is the random Brownian force.

Requiring that the particle reach the Maxwell--Boltzmann equilibrium,
$\langle v_i^2\rangle = k_B T/m_p$, fixes the second‐order statistics of
$\mathbf F_B$:
\begin{equation}
    \bigl\langle F_{B,i}(t)\,F_{B,j}(t') \bigr\rangle
      = 2\,\frac{m_p k_B T}{\tau_p}\,
        \delta_{ij}\,\delta(t-t')
  \label{eq:fd}
\end{equation}
(white noise, zero mean).  Dividing by $m_p^2$ gives the acceleration–noise
covariance
\[
  \bigl\langle a_{B,i}(t)\,a_{B,j}(t') \bigr\rangle
    = 2\,\frac{k_B T}{m_p\tau_p}\,
      \delta_{ij}\,\delta(t-t').
\]

Integrating (\ref{eq:langevin}) from $t$ to $t+\Delta t$,
\[
  \Delta\mathbf v
    = -\frac{\mathbf v}{\tau_p}\,\Delta t
      + \int_{t}^{t+\Delta t}\!\mathbf a_B(t')\,dt'.
\]
For white noise the time integral is normally distributed with
\[
  \bigl\langle\Delta\mathbf v_{\text{Brown}}\bigr\rangle = \mathbf 0,
  \qquad
  \text{Var}\!\left(\Delta v_{\text{Brown},i}\right)
      = 2\,\frac{k_B T}{m_p\tau_p}\,\Delta t.
\]
Hence one may write the stochastic velocity increment as
\begin{equation}
    \Delta\mathbf v_{\text{Brown}}
      = \sqrt{2\,\frac{k_B T}{m_p\tau_p}}\,
        \sqrt{\Delta t}\;
        \hat{\mathbf n}_p
  \label{eq:deltav}
\end{equation}
with $\hat{\mathbf n}_p$ a vector of independent, standard Gaussian deviates.

Many particle solvers apply a constant acceleration within the same
timestep:
\[
  \mathbf a_p^{\text{Brown}}
     = \frac{\Delta\mathbf v_{\text{Brown}}}{\Delta t}.
\]
Substituting (\ref{eq:deltav}) yields
\begin{equation}
   \mathbf a_p^{\text{Brown}}
      = \hat{\mathbf n}_p
        \sqrt{\frac{2}{\tau_p\,\Delta t}\;
               \frac{k_B T}{m_p}}
  \label{eq:abrown}
\end{equation}
which equals \Eq{eq:brown}. The validity of this equation is demonstrated in the following subsection.

\subsection{Validation of expression for Brownian motion}
The diffusion coefficient due to Brownian motion is given by the Einstein relation; $D=k_BT/\gamma$, where $\gamma$ is the friction coefficient.
In the free molecular regime, the friction coefficient can be characterized by the particle response time as $\gamma=m/\tau_p$, such that 
\EQ
\label{eq:diffusion}
D=\tau_p k_B T/m.
\EN
The mean particle displacement is then given by
\EQ
\sigma^2=2 n_{\rm dims}Dt,
\EN
where $\sigma^2=\langle(x-\langle x\rangle)^2\rangle$ and angle brackets represent averaging and $n_{\rm dims}$ is the number of dimensions in which the particle can move.
\begin{figure}[t!]\begin{center}
\includegraphics[width=\columnwidth]{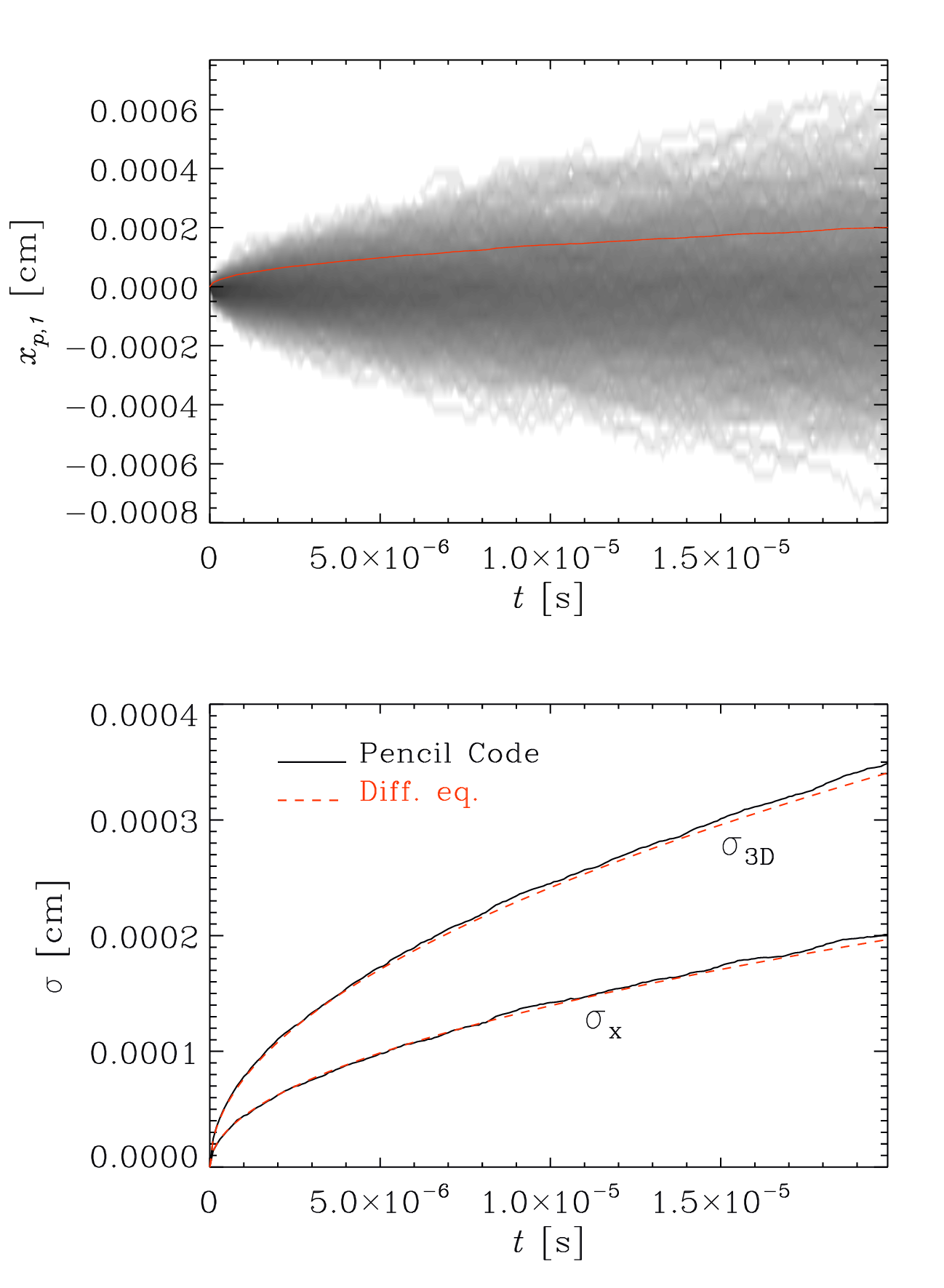}
\end{center}\caption[]{
Heat map of particle position as function of time (upper panel) and mean particle displacement (lower panel). In the upper panel, only 100 of the 1000 test particles are shown.
}\label{show_brownian}\end{figure}

To test the validity of \Eqs{eq:brown}{eq:vrel} and the corresponding implementation in the code, 1000 particles with diameter of $10\nm$ are tracked in a quiescent fluid at $750\K$.
The $x$-position in the domain as a function of time for each of the particles is shown in the upper panel of \Fig{show_brownian}. The red solid line in the same panel represent the mean particle displacement, $\sigma$. In the lower panel of \Fig{show_brownian}, the mean particle displacement from the simulation (solid black lines) is compared with the mean particle displacement as expected from the diffusion equation given by \Eq{eq:diffusion} (dashed red line).
The lower pair of lines corresponds to the full three-dimensional displacement, while the lower pair is for the displacement in the $x$-direction (as visualized in the upper panel).

\section{Condensation}
\label{app:condensation}
\subsection{Molecular/kinematic regime}
In the kinematic regime (particles are much smaller than the mean free path
of the molecules, the effluent flux, which is the rate of molecular collisions per unit surface area, is then given by
\EQ
\mathcal{F}=\Nu/4,
\EN
when $N$ is the concentration of condensing molecules and $u$
is the mean velocity of the gas molecules, which can be found from
the mean of the Maxwell-Boltzmann distribution:
\EQ
u=\sqrt{8k_BT/(\pi m_{\SiOO})},
\EN
where $k_B$ is Boltzmann's constant, $T$ is fluid temperature and $m_{\SiOO}=60m_u$ is
the molecular mass of SiO$_2$ and $m_u=1.67\times 10^{-27}$ kg is the atomic mass constant.
Hence, the units of $\mathcal{F}$ is mol/m$^2$/s. The total molar rate of
condensation on the surface of a particle with radius $r$ is then given by
\EQ
F=4\pi r^2\mathcal{F}.
\EN
From this it is clear that the volumetric change of the particle becomes
\EQ
\frac{dV}{dt}=F M_{\SiOO}/\rho_{\ms},
\EN
where $M_{\SiOO}=60$~kg/kmol is the molar mass of SiO$_2$ and
$\rho_{\ms} = 2196$~kg/m$_3$ is the true density of a
silica fume particle.
When $N=C_{\SiOO}-C_{\rm sat}$ is the molar concentration of the condensing gas,
$C_{\SiOO}$ is the molar concentration of SiO$_2$ gas, $C_{\rm sat}$ is
saturation concentration of the same gas we know that for a spherical particle
$dV=4\pi r^2dr$, such that
\EQA
\label{radius_evolution}
\frac{dr}{dt}=G
&=&\frac{dV}{dt}\frac{1}{4\pi r^2}
=\mathcal{F}4\pi r^2 \frac{M_{\SiOO}}{4\pi r^2\rho_{\ms}} \nonumber \\
&=&(C_{\SiOO}-C_{\rm sat}) \sqrt{8k_BT/(\pi m_{\SiOO})}\frac{M_{\SiOO}}{4\rho_{\ms}} \nonumber \\
&=&A(C_{\SiOO}-C_{\rm sat}) \sqrt{T},
\ENA
where $A=\sqrt{\frac{8k_B}{\pi m_{\SiOO}}}\frac{M_{\SiOO}}{4\rho_{\ms}}$.

\subsection{Continuum regime}
\begin{figure}[t!]\begin{center}
\includegraphics[width=\columnwidth]{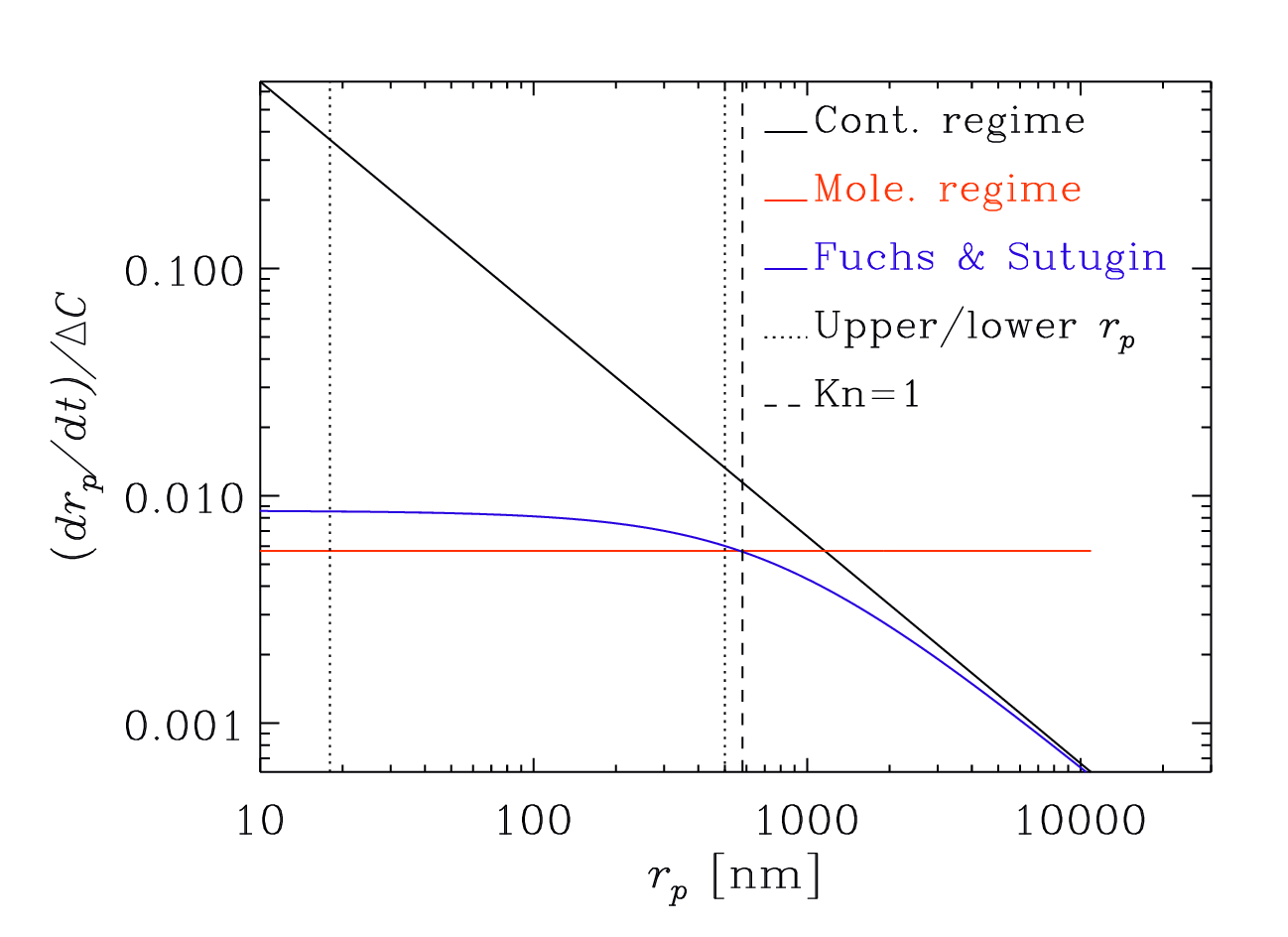}
\end{center}\caption[]{
Particle condensation rates for continuous approximation (black line) and molecular approximation (red line). The Fuchs and Sutugin approximation (blue line) recovers the two regimes relatively well.
}\label{cond_rate}\end{figure}
In the continuum regime, i.e. when the particles are much larger than the mean free path ($\Kn \gg 1$), the radius evolution is given by \cite{Friedlander2000}
\EQA
\label{radius_evolution_cont}
\frac{dr}{dt}=\frac{2D_\SiOO (C_{\SiOO}-C_{\rm sat})M_{\SiOO}}{2r_p\rho_{\ms}}.
\ENA
This is currently not implemented in the simulations below.

Fuchs \& Sutugin proposed the following interpolation formula
\EQA
\label{radius_evolution_cont_inter}
\frac{dr}{dt}&=&\frac{2D_\SiOO (C_{\SiOO}-C_{\rm sat})M_{\SiOO}}{2r_p\rho_{\ms}}
\nonumber \\
&\times&
\left[ 
\frac{1+\Kn}{1+1.71\Kn+1.333\Kn^2}
\right].
\ENA

Assuming air at 1 bar and 2000K, the three above radial evolution equations give the result presented in \Fig{cond_rate}.

\section{Particle size distribution}
The volume-based particle size distribution is defined as
\EQ
f_V(R_i)dr=\frac{1}{V_{\rm total}}\sum_{p=1}^{N_p}V_i(r_p)w.
\EN
Here, $dr=(R_1-R_{\rm max})/N_{\rm bins}$ is the width of each particle bin, $R_i=R_1+(i-1)dr$ is the radius of bin $i$, $N_{\rm bins}$ is the number of bins, $N_p$ is the total number of particle swarms,
\[ V_i(r)=\left\{ \begin{array}
{r@{\quad:\quad}l}
     V(r)& V(R_i)\leq V(r) \le V(R_i+dr)  \\
     0& \mbox{elsewhere} 
\end{array} \right.
\]
is equal to the particle volume if the particle is a member of bin $i$, otherwise it is zero. For experimental measurements, the weight $w$ is unity, while for numerical simulations with the swarm approach, $w=\nsw_p=dV$.
Furthermore, the particle volume is given by
\EQ
V(r)=\frac{4\pi}{3}r^3,
\EN
such that the total volume of all particles become
\EQ
V_{\rm total}=\sum_{p=1}^{N_p}V(r_p)\nsw_p,
\EN
when $\nsw_p$ is the number density of physical particles within swarm $p$.
This ensures that 
\EQ
\sum_{i=1}^{N_{\rm bins}} f_V(R_i)dr=1.
\EN

\hfill(\today)

\vspace{1cm}
\bibliography{references}

\end{document}